\documentclass{jfm}
\usepackage{graphicx}
\usepackage{epstopdf, epsfig}
\usepackage[]{psfrag}
\usepackage{xcolor}
\usepackage{subfig}
\usepackage{tikz}
\usetikzlibrary{decorations.pathreplacing}

\newcommand{\SFnew}[3][]{
  \ensuremath{S_{#3} (s_{#1};u_{#2}) } }
\newcommand{\SFTnew}[3][]{
  \ensuremath{S^T_{#3} (s_{#1};u_{#2}) } }
\newcommand{\var}[1][]{
  \ensuremath{\langle u_{#1}^{\prime 2}\rangle } }
\newcommand\encircle[1]{%
  \tikz[baseline=(X.base)] 
    \node (X) [draw, shape=circle, inner sep=-0.3pt] {\strut #1};}
\newcommand{\upperRomannumeral}[1]{\uppercase\expandafter{\romannumeral#1}}

\newcommand{\rev}[1]{\textcolor{black}{#1}}
  
\definecolor{myblue}{RGB}{25,44,156}
\definecolor{myred}{RGB}{255,37,37}

\shorttitle{Statistics of turbulence in the energy-containing range}
\shortauthor{Krug et al.}

\title{Statistics of turbulence in the energy-containing range of Taylor-Couette compared to canonical wall-bounded flows}

\author{Dominik Krug\aff{1}
  \corresp{\email{dominik.krug@unimelb.edu.au}},
  Xiang I.A. Yang\aff{2},
  Charitha M. de Silva\aff{1},
  Rodolfo Ostilla-M\'{o}nico\aff{3,4},
  Roberto Verzicco\aff{5,6},
  Ivan Marusic\aff{1},
 \and Detlef Lohse\aff{6,7}}

\affiliation{
$^1$Department of Mechanical Engineering, University of Melbourne, Melbourne, Australia\\
$^2$ Center for Turbulence Research, Stanford University, Stanford, 94305, USA\\
$^3$School of Engineering and Applied Sciences and Kavli Institute for Bionano Science and
Technology, Harvard University, Cambridge, MA 02138, USA\\
$^4$Kavli Institute for Theoretical Physics, Kohn Hall, University of California, Santa Barbara, CA 93106-4030, USA \\
$^5$Dipartimento di Ingegneria Industriale, University of Rome `Tor Vergata', Via del
Politecnico 1 Roma 00133, Italy\\
$^6$Physics of Fluids Group and Twente Max Planck Center, 
Department of Science and Technology, Mesa+ Institute,
 and J.M. Burgers Center for Fluid Dynamics, University of Twente, P.O Box 217, 7500 AE Enschede,  The Netherlands \\
 $^7$Max Planck Institute for Dynamics and Self-Organization, 37077 G\"ottingen, Germany}

\begin{document}

\maketitle

\begin{abstract}
Considering structure functions of the streamwise velocity component in a framework akin to the extended self-similarity hypothesis (ESS), de Silva \textit{et al.} (\textit{J. Fluid Mech.}, vol. 823,2017, pp. 498-510) observed that remarkably the \textit{large-scale} (energy-containing range) statistics in canonical wall bounded flows exhibit universal behaviour. In the present study, we extend this universality, which was seen to encompass also flows at moderate Reynolds number, to Taylor-Couette flow. In doing so, we find that also the transversal structure function of the  spanwise velocity component exhibits the same universal behaviour across all flow types considered. We further demonstrate that these observations are consistent with predictions developed based on an attached-eddy hypothesis. These considerations also yield a possible explanation for the efficacy of the ESS framework by showing that it relaxes the self-similarity assumption for the attached eddy contributions.
By taking the effect of streamwise alignment into account, the attached eddy model predicts different behaviour for structure functions in the streamwise and in the spanwise directions and that this effect cancels in the ESS-framework --- both consistent with the data. Moreover, it is demonstrated here that also the additive constants, which were previously believed to be flow dependent, are indeed universal at least in turbulent boundary layers and pipe flow where high-Reynolds number data are currently available.

\end{abstract}


\section{Introduction}

Structure functions, which characterize the turbulent velocity field in terms of the velocity difference between two points with varying separations, play a vital role in turbulence theory ever since \citet{Kolmogorov1941} formulated his famous `K41' scaling law in the inertial sub-range (ISR). Later on, universal deviations from the K41 scaling were experimentally found in the ISR scaling of velocity structure functions of various orders \citep[e.g.][]{Anselmet1984,Frisch1995}. To firmly establish the  universality of the ISR statistics, the so-called extended self-similarity (ESS) hypothesis of \citet{Benzi1993,Benzi1995} played a central role \citep{Arneodo1996,Belin1996}.  In this framework,  instead of evaluating the scaling of the structure functions  directly as a function of distance, the relative scaling exponent is sought by plotting one structure function against another one of different order on log-log scales. Doing so has been shown to greatly increase the scaling range, rendering the ISR scalings accessible even at only moderate Reynolds numbers. \rev{An application of this concept to channel flow was reported by \citet{Toschi1999}}.

While these seminal findings date back a quarter of a century, a more recent string of research is concerned with the scaling of structure functions in the energy containing range (ECR) in wall-bounded flows. Related studies were initiated by \citet{Davidson2006b,Davidson2006} who showed that the spatial equivalent to the  $k^{-1}$-spectral scaling (where $k$ is the streamwise wavenumber) of the streamwise kinetic energy \citep{Nickels2005}, is a logarithmic behaviour of the second-order structure function. \citet{Davidson2006b} argued that the latter is more readily discernible in the data since it is not subject to aliasing effects present in the spectral domain.
Later on, \citet{deSilva2015scaling} provided evidence for logarithmic scaling also for higher order structure functions and gave a theoretical underpinning for this observation  based on an attached eddy framework and the results of \citet{Woodcock2015}. Specifically, \citet{deSilva2015scaling} investigated even-order streamwise longitudinal structure functions 
\begin{equation}
S_{p}(s_x;u_x) \equiv \langle [\Delta u_x^+(s_x)]^{2p}\rangle^{1/p}.
\end{equation}
Here, $\langle \cdot \rangle$ is an ensemble average, $2p$ denotes the order, $u_x$ the streamwise velocity component and
\begin{equation}
\Delta u_i(s_i) =u_i(\mathbf{x})- u_i(\mathbf{x}+\mathbf{i}s_i)
\end{equation}
is  the velocity increment of a given velocity component $u_i$ between two points $\mathbf{x}$ and $\mathbf{x}+\mathbf{i}s_i$, where $s_i$ is the separation distance along the unit vector $\mathbf{i}$, which points in the direction of the $u_i$-component. Throughout this paper, we indicate normalizations by inner scales $U_\tau$ (mean friction velocity) and $\nu$ (kinematic viscosity) by superscript $+$.
\citet{deSilva2015scaling} found that at a distance $y_0$ from the wall and for the range $y_0<s_x\ll \delta$ (i.e. the ECR), $S_{p}(s_x;u_x)$  scales according to
\begin{equation}
S_{p}(s_x;u_x)  = E_p +D_p \ln \frac{s_x}{y_0},
\label{eq:logSF}
\end{equation}
where $\delta$ is an outer length scale of the flow and  $E_p,D_p$ are constants. However, their work highlighted that  high Reynolds numbers of $Re_{\tau}=\delta U_\tau/\nu \sim O(10^4)$ are required to clearly observe the scaling according to (\ref{eq:logSF}) directly. Moreover, comparison of single-point statistics in the ECR in turbulent boundary layers and channel flows \citep{Sillero2013} and the analysis of the second-order structure function in pipe flows \citep{Chung2015} suggest that differences with regard to (\ref{eq:logSF}) exist in different flow geometries.
However, borrowing inspiration from the original ESS analysis of \citet{Benzi1993,Benzi1995}, \citet{deSilva2017} were recently able to demonstrate universality for the ECR scales in wall-bounded flows. In particular, they showed that when evaluating $\SFnew[x]{x}{p}$ with respect to a reference structure function $S_m(s_x;u_x) \equiv \langle (\Delta u_x^+)^{2m}\rangle^{1/m}$ of arbitrary order $2m$, the `ESS-form' of (\ref{eq:logSF}), given by
\begin{equation}
\SFnew[x]{x}{p} = \frac{D_p}{D_m}S_m(s_x;u_x) + \underbrace{E_p - \frac{D_p}{D_m}E_m}_{E^*_{p,m}},
\label{eq:ESS}
\end{equation}
holds over a larger range of wall distances and at significantly lower $Re_{\tau}$ than the direct representation (\ref{eq:logSF}). It was further established that the ratios $D_p/D_m$ exhibit universality for canonical wall-bounded flows, i.e. for flat plate turbulent boundary layers (TBL), channel flow (CH) and pipe flow. This universality was seen to also comprise the transversal structure functions of $u_x$ defined by 
\begin{equation}
\SFTnew[j]{x}{p} = \langle \left[\Delta^T u_x^+(s_z)\right]^{2p}\rangle^{1/p},
\end{equation}
with
\begin{equation}
\Delta^T u_i (s_j) =u_i(\mathbf{x})- u_i(\mathbf{x}+\mathbf{j}s_j)
\end{equation}
 where $\mathbf{j}$ denotes a unit vector perpendicular to $\mathbf{i}$ in the wall parallel plane and $s_j$ is a distance along $\mathbf{j}$. The reference structure function is denoted as $S_m^T(s_j;u_x)$ for this case.
 
The opportunity created by these results lies in the fact that the formulation of (\ref{eq:ESS}) allows us to study the scaling relations pertinent to (\ref{eq:logSF}) at Reynolds numbers that are accessible via direct numerical simulations (DNS). We will exploit this benefit in the present paper to investigate the ECR scalings for Taylor-Couette (TC) flow in the gap between two coaxial independently rotating cylinders. The rich flow physics of this problem are well studied in many aspects and we refer to the recent reviews by \citet{Fardin2014}, which focusses on flow patterns emerging at moderate Reynolds numbers, and in particular to the one by \citet{Grossmann2016} for a comprehensive overview on high Reynolds number dynamics. Our interest here lies in the highly turbulent state, the so-called ultimate regime, in which the bulk flow as well as the boundary layers are turbulent \citep{Grossmann2011,Huisman2013,Ostilla2014,Grossmann2016}.  For this case, \citet{Huisman2013} recently confirmed the existence of a logarithmic region within the boundary layers and measured a von K\'arm\'an constant $\kappa \approx 0.4$ closely matching values established in other wall-bounded flows \citep{Nagib2008,Marusic2013}. This agreement may appear surprising when noting that unlike in pipe flow, the streamwise curvature in TC flow gives rise to an additional centrifugal instability of the flow, which manifests itself in the presence of Taylor rolls \citep{Taylor1923}. Further, these large-scale coherent structures extend across the entire gap width  and significantly modify the mean flow. Their presence is not limited to the transitional regime, where the boundary layers are laminar and only the bulk flow is turbulent, but also affects the fully turbulent, or ultimate regime  \citep{Ostilla2014,Ostilla2016} up  to very high Reynolds numbers \citep{Huisman2014}. In this context, it appears interesting to check whether the similarity between TC and other wall-bounded flows observed so far also extends to two-point statistics. In particular our focus in this paper is to investigate \textit{(i)} whether a scaling according to the ESS form (\ref{eq:ESS}) also exists in turbulent TC flow and \textit{(ii)} whether turbulent TC flow adheres to the same universality observed in other geometries. 

However, before we proceed to address these questions in \S\ref{sec:TCres}, we will extend the analysis of \citet{deSilva2017} in some aspects. Specifically, we will derive the scaling of higher order even structure functions according to (\ref{eq:logSF}) using an attached-eddy argument in the spirit of the hierarchical-random-additive-process \citep{Meneveau2013,Yang2016jfm}  in \S\ref{sec:AE}. This analysis also suggests an explanation as to why the ESS form (\ref{eq:ESS}) extends the scaling range by demonstrating that it relaxes the strong self-similarity assumption required for (\ref{eq:logSF}). 
Another issue that has not yet received any attention  is the behaviour of the additive constant in the ESS form, which we label $E^*_{p,m}$ as indicated in (\ref{eq:ESS}). As our data presented in \S\ref{sec:ADD} reveal, $E^*_{p,m}$ also exhibits a certain degree of universality when plotted as a function of the variance $\langle {u_x^\prime }^2\rangle$, where the velocity fluctuation $u_i^\prime \equiv u_i -U_i$ and the mean $U_i \equiv \langle u_i \rangle$. Additionally, we give a short summary of all the datasets employed in this study in \S\ref{sec:DATA}, consider higher-order structure functions in \S\ref{sec:TCres} and present our conclusions in \S\ref{sec:CONC}.


\section{Analysis within an attached eddy framework} \label{sec:AE}

\subsection{Calculation of the structure functions}
\begin{figure} 
\psfrag{a}[c][c][1]{$(a)$}
\psfrag{b}[c][c][1]{$(b)$}
\psfrag{x}[c][c][0.8]{$x$}
\psfrag{y}[c][c][0.8]{$y$}
\psfrag{z}[c][c][0.8]{$z$}
\psfrag{sx}[c][c][1]{$s_x$}
\psfrag{sz}[c][c][1]{$s_z$}

\psfrag{y0}[c][c][1]{$y_0$}
\psfrag{1}[c][c][0.8]{\upperRomannumeral{1}}
\psfrag{2}[c][c][0.8]{\upperRomannumeral{2}}
\psfrag{3}[c][c][0.8]{\upperRomannumeral{3}}
\psfrag{4}[c][c][0.8]{\upperRomannumeral{4}}
\psfrag{5}[c][c][0.8]{\upperRomannumeral{5}}
\psfrag{6}[c][c][0.8]{\upperRomannumeral{6}}

\psfrag{e}[c][c][0.8]{$O(10\delta)$}
\psfrag{s}[l][l][0.8]{\textcolor{myblue}{low speed}}
\psfrag{hs}[l][l][0.8]{\textcolor{myred}{high speed}}
\psfrag{o}[c][c][0.8]{\encircle{1}}
\psfrag{t}[c][c][0.8]{\encircle{2}}

\psfrag{phi}[c][c][0.8]{$\phi$}
\psfrag{uphi}[c][c][0.8]{$u_i(\phi)$}
\psfrag{usphi}[c][c][0.8][-12]{$u_i(\phi+\delta\phi)$}
\psfrag{sphi}[c][c][0.8]{$s_{\phi}$}
\psfrag{ri}[c][c][0.8]{$r_i$}
\psfrag{ro}[c][c][0.8]{$r_o$}
\psfrag{omi}[c][c][0.8]{$\omega_i$}
\psfrag{d}[c][c][0.8]{$d$}
\psfrag{meanuphi}[l][l][0.8]{$\widetilde{U}_{\phi}(y)$}
\centering
\vspace{0.5cm}
\subfloat{\includegraphics[scale = 0.6]{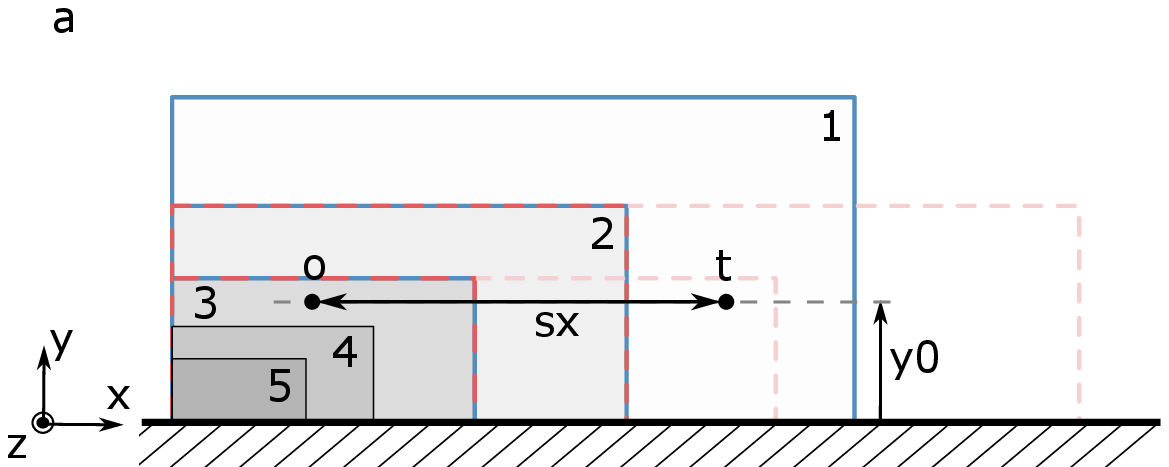}}\hfill
\subfloat{\raisebox{0.0cm}{\includegraphics[scale = 0.6]{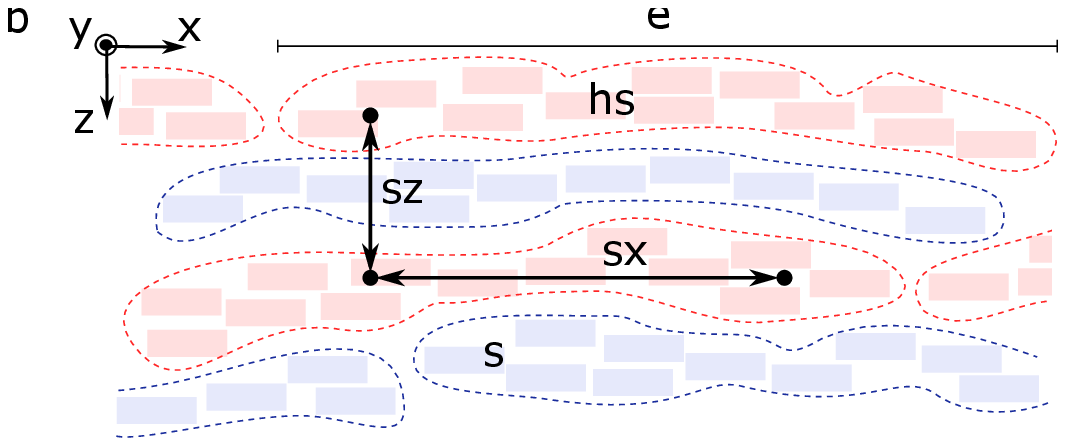}}}\hspace{0.0cm}
\caption{\label{fig:sketchAE} (a) Side-view sketch of the random additives at different hierarchies I-V with an arbitrarily chosen scaling factor of 1.5; hierarchies that contribute to single-point statistics at point \protect\encircle{1} are marked blue, those contributing to the difference between \protect\encircle{1} and \protect\encircle{2} in red. (b) Top-view illustration of the effect of long streamwise structures on the structure functions in different directions. Individual $a$-eddy contributions at a single hierarchy are represented as boxes and coloured red (blue) if they belong to a \rev{high (low)} speed structure. The bounds of the large scale structures (dashed lines) may be interpreted as $b$-eddies. 
 }
\end{figure}

We consider a simplified version of the attached-eddy model \citep[e.g.][]{Townsend1976,Perry1982mechanism,Perry1986}, by adopting a hierarchical-random-additive-process (HRAP) proposed by \citet{Meneveau2013} and \citet{Yang2016}. Here, velocity fluctuations at a given point in the flow are modelled as a result of an instantaneous superposition of attached-eddy contributions. For simplicity, we will restrict the discussion to the streamwise velocity component initially, where the fluctuating part is given by the sum
\begin{equation}
u^+_x(y)=\sum_{l=1}^{N(y)}a_l.
\label{eq:add}
\end{equation}
The random additives $a_l$ signify contributions from single, spatially self similar  wall-attached `eddy' structures and the summation is over different hierarchies of eddies, with small $l$ representing the large eddies. 
\rev{Assuming the eddy population density to be inversely proportional to $y$ as originally  proposed by \citet{Townsend1976}, the number  of summands  contributing at $y$ is related to the distance off the wall by 
\begin{equation}
N(y) \propto \int_y^{\delta}\frac{1}{y}dy = \ln(\delta/y),
\label{eq:Ny}
\end{equation}
 where $\delta$ is an outer length scale.} A schematic representation of this simple conceptual model is provided in figure \ref{fig:sketchAE}a.
Following \citet{Kolmogorov1962}, we make the simplifying assumption that the random additives follow a Gaussian distribution with zero mean and variance $\sigma$ denoted by $\mathcal{N}(0,\sigma^2)$. The mean is zero since we only consider velocity fluctuations. Indeed, the arrival of an attached eddy can be assumed to follow roughly a Poisson process \citep[e.g.][]{Woodcock2015} and the velocity fluctuations are results of the superposition of the eddy induced velocity fields such that the Gaussian approximation appears justified. In this case, we can restate (\ref{eq:add}) as
\begin{equation}
u^+_x(y) = \mathcal{N}\left(0,\sum_{l=1}^{N(y)}a_l^2 \right)=\mathcal{N}\left(0,N(y) \langle a^2\rangle \right),
\label{eq:momGauss}
\end{equation}
where the second step exploits the self-similarity of the eddies. From this and (\ref{eq:Ny}), it follows from the properties of the Gaussian distribution  that for even-order moments
\begin{equation}
\left<(u^+_x(y))^{2p}\right>^{1/p}
=\hat{B}_p-\left[(2p-1)!!\right]^{1/p}\hat{A}_1\ln(y/\delta),
\label{eq:momlog}
\end{equation}
 which recovers the result of \citet{Meneveau2013} with the `Townsend-Perry' constant $\hat{A}_1$ denoting the slope of the log law for the variance $\left<(u^+_x(y))^{2}\right>$ (for which the double factorial $(2-1)!!=1$) and $\hat{B}_p$ is an additive constant. Note that the hat is used here to differentiate from the actual experimentally obtained constants since (\ref{eq:add}) is only an approximation of the actual velocity field.
 
Taking the difference between velocity fluctuations at two points cancels out the contributions of common eddies as they are assumed to affect both points equally. Therefore  only eddies that affect each point individually  remain in this case. We assume that a typical attached eddy has an aspect ratio $R_x$ given by its streamwise extent over its height. Then for a given streamwise separation $s_x$ only eddies smaller than $y_d = s_x/R_x$ contribute to the difference, i.e. large eddies that are shared between the two points  (corresponding to small $l$) drop out as illustrated in figure \ref{fig:sketchAE}a. As a result
\begin{equation}
u^+_x(x,y)-u^+_x(x+s_x,y)= \Delta u^+_x(s_x) =\sum_{l=N(y_d)}^{N(y)}a_l - \sum_{l=N(y_d)}^{N(y)}a_l^\prime,
\label{eq:du-HRAP}
\end{equation}
where the prime denotes independent eddy contributions at a second point $x+s_x$. The same Gaussian approximation leading to (\ref{eq:momGauss}) yields in this case 
\begin{equation}
\Delta u^+_x(s_x)=\mathcal{N}\left(0,\sum_{l=N(y_d)}^{N(y)}(a_l^2+{a_l^\prime}^2)\right).
\label{eq:du-Norm}
\end{equation}
Assuming $a_l$ and $a_l^\prime$, which relate to the same hierarchy at different positions, to be identically distributed based on their self-similarity and again using (\ref{eq:Ny}), we arrive at
\begin{equation}
\Delta u^+_x(s_x)=\mathcal{N}\left(0,2\left[N(y)-N(y_d)\right]\left<a^2\right>\right)=\mathcal{N}\left(0,2\ln(y_d/y)\left<a^2\right>\right).
\label{eq:du-NormSS}
\end{equation}
\rev{We would like to caution the reader that the exact cancelling of 'shared' eddies (leading to (\ref{eq:du-HRAP})) or the complete independence of eddies at different locations (leading to (\ref{eq:du-Norm})) have to be viewed as first order approximations and cannot be expected to hold exactly in practice \cite[see also][in this regard]{Davidson2004,Davidson2006b}. We present in \S\ref{sec:refine} how the framework can be refined to account for streamwise alignment, i.e. incomplete decorrelation, of eddies.}

From this result, the scaling relation  for even order structure functions  is obtained analogous to (\ref{eq:momlog}) as
\begin{equation}
\SFnew[x]{x}{p}
=\left[(2p-1)!!\right]^{1/p}2\left<a^2\right>\ln(y_d/y)
=\left[(2p-1)!!\right]^{1/p}2\hat{A}_1 \ln(s_x/y) +\hat{E}_p,
\label{eq:du-SS}
\end{equation}
where $\hat{E}_p$ is a constant. The logarithmic scaling of higher order structure functions predicted here was recently observed by \citet{deSilva2015scaling}. Typically, the actual  slopes in the semi-logarithmic plots $\SFnew[x]{x}{p}$ vs. $\ln(s_x/y)$ will be smaller than the ones predicted in (\ref{eq:momlog}) and (\ref{eq:du-SS}) as wall-bounded flows are known to be sub-Gaussian \citep{Jimenez1998,Meneveau2013}. We will test this for the ECR slopes in \S\ref{sec:TCresD}.

\subsection{Extended self-similarity for structure functions}
In the following, we will show that the ESS-framework relaxes the strong self-similarity assumption necessary to derive (\ref{eq:du-NormSS}) using an approach similar to \citet{Yang2016}. This is useful as in practice self-similarity may be broken by large-scale effects as well as viscous (small-scale) effects such that the scaling according to (\ref{eq:du-SS}) may not be discernible. Without the strong self-similarity assumption we have
\begin{equation}
\SFnew[x]{x}{p}=\left[(2p-1)!!\right]^{1/p}\left(2\sum_{l=N(y_d)}^{N(y)}\left<a_l^2\right>\right),
\label{du-ESS}
\end{equation}
but now $\left<a_l^2\right>\neq \left<a_k^2\right>$ if $l \neq k$.
Although no scaling according to (\ref{eq:du-SS}) can be obtained directly in this case, forming the ratio in the spirit of ESS gives 
\begin{equation}
\frac{\SFnew[x]{x}{p}}{\SFnew[x]{x}{m}}
=\frac{\left[(2p-1)!!\right]^{1/p}\left(2\sum_{l=N(y_d)}^{N(y)}\left<a_l^2\right>\right)}{\left[(2m-1)!!\right]^{1/m}\left(2\sum_{l=N(y_d)}^{N(y)}\left<a_l^2\right>\right)}=\frac{\left[(2p-1)!!\right]^{1/p}}{\left[(2m-1)!!\right]^{1/m}}.
\label{eq:du2du-ESS}
\end{equation}
This shows that even if the self-similarity of the eddies is broken, ESS  establishes a linear relationship between $\SFnew[x]{x}{p}$ and $\SFnew[x]{x}{m}$. 
We note that although (\ref{eq:du2du-ESS}) is trivial under the Gaussian assumption employed here, it is still valuable in demonstrating the functional principle of ESS. In general, $\Delta u^+_x(s_x)$ is non-Gaussian, which e.g. manifests itself in the presence of a non-zero additive constant in (\ref{eq:ESS}), such that this result is non-trivial.
Formally, (\ref{eq:du2du-ESS}) holds for all $y_0$ and $s_x$. However, at small-scales $\eta \ll s_x\ll y_0$, where the Kolmogorov scale $\eta$ is the length scale of the smallest, dissipative structures in the flow, the  scaling of the inertial sub-range will prevail. Unlike the logarithmic relationship (\ref{eq:logSF}) for the ECR, the ISR scaling has the form of a power-law. 
Hence, instead of  (\ref{eq:du2du-ESS}),  the classical ESS in the inertial range \citep{Benzi1993,Benzi1995}  manifests itself  as the ratio of the logarithms of the structure functions according to
\begin{equation}
\frac{\ln\left[\SFnew[x]{x}{p}\right]}{\ln\left[\SFnew[x]{x}{m} \right]}=\textrm{const.}
\end{equation}
This serves to highlight the differences between the original ESS hypothesis and the related framework introduced in \citet{deSilva2017} and employed here. We will not pursue ISR scaling further in the following.

\subsection{Extension to the transversal direction and the spanwise velocity component} \label{sec:refine}
It is easy to see that the above arguments can readily be extended to transversal structure functions. To this end, we define the spanwise aspect ratio $R_z$ as the ratio of the spanwise extent of the attached eddies to their height. With this definition, we have $y_d = s_z/R_z$ as the height of the largest eddy contributing to the velocity difference between two points separated by $s_z$ in the spanwise direction, leading to the same result as (\ref{eq:du-SS}). However, since $R_x/R_z>1$, we expect the logarithmic scaling range at smaller separations in the spanwise direction than in the streamwise direction.

It is noteworthy that under the assumptions leading to (\ref{eq:add}), the slope in (\ref{eq:du-SS}) remains unchanged between the streamwise and spanwise directions regardless of $R_x/R_z$ since the aspect ratio only affects the additive constant.
A more realistic prediction can be obtained, however, when experimental evidence for the existence of very long  structures \citep{Hutchins2007}, whose streamwise extent far exceeds the logarithmic scaling region, is taken into account. Similar to \citet{Tomkins2003}, it may be assumed that in this case eddy contributions remain correlated over large streamwise distances as is illustrated in figure \ref{fig:sketchAE}b. Hence, the assumption of independence between eddy contributions $a_l$ and $a_l^\prime$ for points separated in $x$, which was made in deriving (\ref{eq:du-SS}), is no longer readily applicable.
Mathematically, we model this effect by considering an additional type of eddy contributions  denoted as $b_l$  (and referred to as `$b$-eddies' in the following as opposed to the `$a$-eddies' considered so far). Doing so is in line with e.g. \citet{Perry1995}, who - albeit for different reasons - considered a total of three eddy types. The streamwise length of the structures ($O(10\delta)$) is large compared to the upper bound of the ECR-scaling ($\delta$), such that we  assume that $R_x^b \to \infty$ (superscript $b$ denotes quantities pertaining to $b$-eddies), while $R^b_z$  remains finite and  all other assumptions for $a$-eddies apply. This implies that $N^b(y) = N(y)$ and instead of (\ref{eq:add}) the model velocity is then given to leading order by
\begin{equation}
u^+_x(y)=\sum_{l=1}^{N(y)}a_l +\sum_{l=1}^{N(y)}b_l.
\label{eq:add2}
\end{equation}
  
For structure functions in the streamwise direction, the model prediction (\ref{eq:du-SS}) remains unchanged since $y_d^b \to 0$ for $R_x^b \to \infty$, i.e. the $b$-eddies always affect both points and therefore do not contribute to the velocity difference.  However, for the spanwise direction using (\ref{eq:add2}) leads to 
\begin{equation}
\Delta u^+_x(s_z) =\sum_{l=N(y_d)}^{N(y)}a_l-\sum_{l=N(y_d)}^{N(y)}a_l^\prime
+\sum_{l=N(y_d)}^{N(y)}b_l-\sum_{l=N(y_d)}^{N(y)}b_l^\prime,
\end{equation}
from which under the assumption of independent Gaussian random processes and following the same arguments leading to (\ref{eq:du-NormSS} - \ref{eq:du-SS}), the logarithmic scaling according to 
\begin{equation}
\SFTnew[z]{x}{p}
=\left[(2p-1)!!\right]^{1/p}2\left<a^2+b^2\right>\ln(s_z/y)+\hat{E}_p
\end{equation} 
is obtained.
Comparing this result to the equivalent for $\SFnew[x]{x}{p}$ in (\ref{eq:du-SS}), it becomes evident that the logarithmic slope in the spanwise direction is predicted to be greater.
  Nevertheless, the ESS approach will cancel this effect in a similar vein to (\ref{eq:du2du-ESS}) even if $b_l \neq b_k$ if $l \neq k$ as is seen from
\begin{equation}
\frac{\SFTnew[z]{x}{p}}{\SFTnew[z]{x}{m}}
=\frac{	\left[(2p-1)!!\right]^{1/p}		2\sum_{l =N(y_d)}^{N(y)}\left(\left<a_l^2\right>+\left<b_l^2\right>\right)}
{		\left[(2m-1)!!\right]^{1/m}		2\sum_{l=N(y_d)}^{N(y)}	\left(\left<a_l^2\right>+\left<b_l^2\right>\right)}=\frac{\left[(2p-1)!!\right]^{1/p}}{\left[(2m-1)!!\right]^{1/m}},
\label{eq:du2du-ESS-trans}
\end{equation}
 such that the prediction for the relative slopes $D_p/D_m$ from the attached eddy model is the same in streamwise and spanwise directions. We will test these implications on the data in \S\ref{sec:TCres}.
 {\color{blue}
}

Moreover, an equivalent ansatz to (\ref{eq:add}) or (\ref{eq:add2}) can also be made for the spanwise velocity component $u_z$ from which it follows that all of the results of this section are also applicable in this case.
Investigations on the spanwise velocity component in wall-bounded flows are generally scarce \citep{Pirozzoli2013,Sillero2013,Stevens2014,Talluru2014} and evidence of an ECR scaling according to (\ref{eq:logSF}) in this case is missing to date. The data necessary for such an investigation will be available from the simulations employed here. We will therefore include a study of the ECR framework applied to $u_z$ in the following. Especially for the TC case, this is of great interest since unlike for the streamwise velocity component, the Taylor rolls directly contribute to $u_z$.

\section{Flow geometries and datasets } \label{sec:DATA}
\subsection{Numerical dataset for TC flow}
\begin{figure} 
\psfrag{a}[c][c][1]{$(b)$}
\psfrag{b}[c][c][1]{$(a)$}
\psfrag{x}[c][c][0.8]{$x$}
\psfrag{y}[c][c][0.8]{$y$}
\psfrag{z}[c][c][0.8]{$z$}
\psfrag{h}[c][c][0.8]{$h$}
\psfrag{y0}[c][c][0.8]{$y_0$}
\psfrag{meanu}[c][c][0.8]{$U_x(y) $}
\psfrag{ux}[c][c][0.8]{$u_i(x)$}
\psfrag{uxsx}[c][c][0.8]{$u_i(x+s_x)$}
\psfrag{sx}[c][c][0.8]{$s_x$}
\psfrag{phi}[c][c][0.8]{$\phi$}
\psfrag{uphi}[c][c][0.8]{$u_i(\phi)$}
\psfrag{usphi}[c][c][0.8][-12]{$u_i(\phi+\delta\phi)$}
\psfrag{sphi}[c][c][0.8]{$s_{\phi}$}
\psfrag{ri}[c][c][0.8]{$r_i$}
\psfrag{ro}[c][c][0.8]{$r_o$}
\psfrag{omi}[c][c][0.8]{$\omega_i$}
\psfrag{d}[c][c][0.8]{$d$}
\psfrag{meanuphi}[l][l][0.8]{$\widetilde{U}_{\phi}(y)$}
\centering
\vspace{0.5cm}
\subfloat{\includegraphics[scale = 0.6]{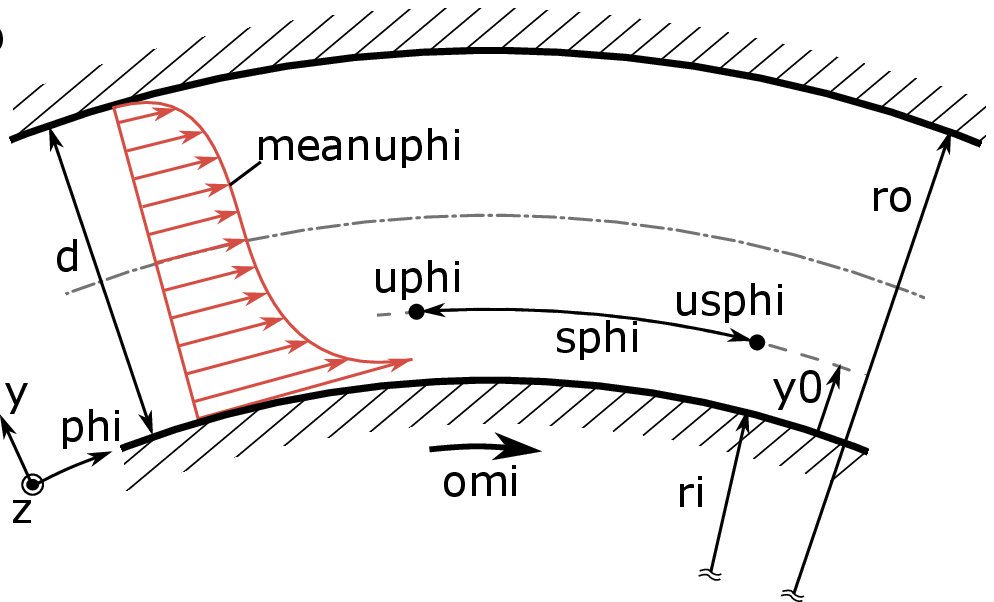}}
\subfloat{\raisebox{0.85cm}{\includegraphics[scale = 0.55]{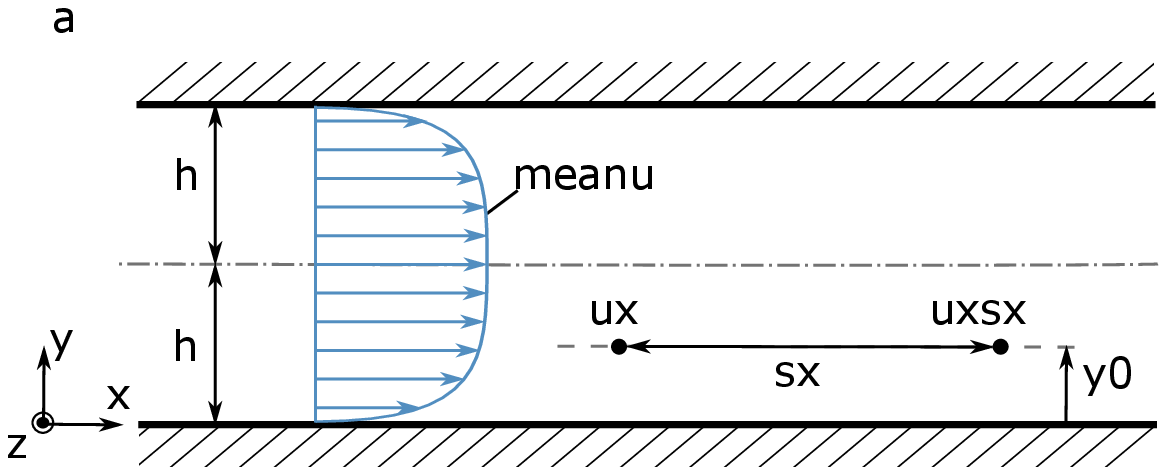}}}\hspace{0.50cm}
\caption{\label{fig:sketch} Sketches of the geometries of Taylor-Couette (a) and channel flow (b). Also shown are the configurations of structure functions along the  streamwise directions.  }
\end{figure}

The geometry of the TC problem is sketched in figure \ref{fig:sketch}a. Our results are based on the simulation labelled `R3' in  \citet{Ostilla2016} and we refer to the original work for additional details on the numerical setup. We restrict the discussion to the case of pure inner-cylinder rotation (with angular velocity $\omega_i$) and a radius ratio $\eta \equiv	r_i/r_o = 0.909$ (subscripts '$i$' and $'o'$ label quantities related to the inner and outer cylinder, respectively). 
The Reynolds number based on half the gap width $d = r_o-r_i$ and the shear stress at the inner cylinder is $Re_{\tau,i} = U_{\tau,i} d/(2\nu)=3920$. In order to be able to afford such a relatively high Reynolds number,  the computational domain is kept small by making use of axial as well as an imposed azimuthal symmetry of order 20, that is only a 1/20th segment of the domain is simulated.  \rev{The simulation employs $N_{\phi}\times N_r \times N_z = 2048\times 1536 \times 3072$ grid points resulting in spatial resolutions at mid gap in inner wall units of $\mathcal{D}^+_{\phi}(r_i+r_o)/2 =12.6$ and $\mathcal{D}_z^+ = 5.1$, where $\mathcal{D}_{\phi}$ and $\mathcal{D}_{z}$ are grid spacings in $\phi$ and $z$, respectively. } Due to its limited axial extent, the box only contains one pair of Taylor rolls, which was however shown to be sufficient to represent the relevant flow statistics \citep{Brauckmann2013,Ostilla2015}. Further, the effect of the number of rolls on the flow statistics has been studied in detail by \citet{Ostilla2016rolls}.
The rolls are oriented along the azimuthal direction (i.e. they contribute predominantly to the wall-normal and axial velocity components) and we will refer to their axial wavelength as $\lambda_{TR}$. As depicted in figure \ref{fig:sketch}a, we define the wall-normal coordinate $y$ as the distance from the inner cylinder, while $z$ and $\phi$ denote the spanwise and azimuthal directions, respectively. We will present results for the boundary layer at the inner cylinder only. Therefore, we define $u_\phi$ as the velocity deficit $u_\phi = \omega_i r_i - \widetilde{u}_\phi$ with $\widetilde{u}_\phi$ denoting the actual azimuthal velocity whose mean profile is sketched in figure \ref{fig:sketch}a. For convenience, we nevertheless refer to $u_\phi$ as the `azimuthal' or `streamwise' velocity component in this paper. Figure \ref{fig:sketch}a further includes a sketch of the definitions for structure functions with separations along the streamwise directions. In the case of TC, the distance $s_\phi$ is measured along the circumference such that $s_{\phi}=\delta \phi(r_i+y)$, where $\delta \phi$ denotes the difference in $\phi$ between the two points. Apart from the spatial averages over homogeneous directions, our results for TC are averaged over 5 independent snapshots in time, which is sufficient for statistical convergence if not stated otherwise.

\subsection{Numerical dataset for channel flow and comparison}

\begin{figure} 
\psfrag{a}[c][c][1]{$(a)$}
\psfrag{b}[c][c][1]{$(b)$}
\psfrag{meanp}[c][c][1]{$U^+_x,\, U^+_{\phi}$}
\psfrag{varp}[c][c][1]{$\langle u_x^{\prime 2}\rangle^+, \, \langle u_{\phi}^{\prime 2}\rangle^+$}
\psfrag{zp}[c][c][1]{$y^+$}
\psfrag{log}[r][r][0.9]{$\propto 2.6\ln(y^+)$}

\psfrag{CH}[l][l][1]{Channel}
\psfrag{TC}[l][l][1]{TC}
\centering
{\includegraphics[scale = 0.75]{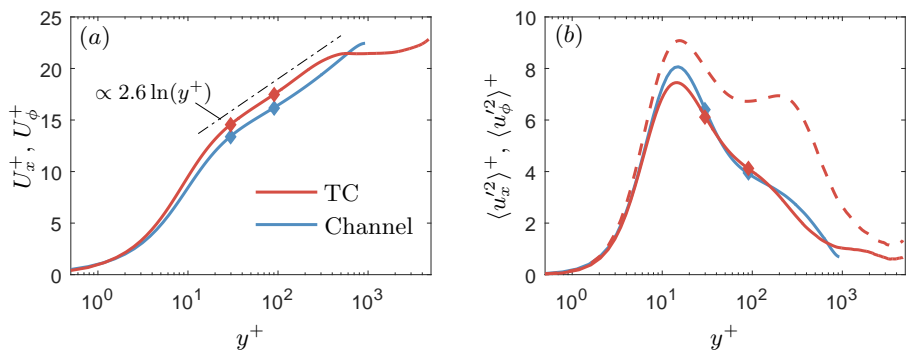}}
\caption{\label{fig:meanvar} Mean (a) and variance (b) of the streamwise (for channel flow) and azimuthal (for TC) velocity component; dashed dotted line in (a) is $U^+\propto 2.6\ln(y^+)$ for reference, red solid line in (b): TC variance based on $y$-dependent mean, dashed line: TC variance based on a conventional mean. Diamonds mark the position for which the structure functions are investigated in \S\ref{sec:TCres}.  }
\end{figure}

For comparison to a canonical wall-bounded flow, we also include results based on the channel simulations of  \citet{delAlamo2004}, where 15 independent snapshots are available. The corresponding geometry  is shown in figure \ref{fig:sketch}b. Even though with $Re_\tau = U_\tau h/\nu \approx 1900$ ($h$ being the half width), the nominal Reynolds number is somewhat lower in this case compared to the TC data, the flows are indeed quite similar in the proximity of the wall as figure \ref{fig:meanvar} shows. In that figure we plot the mean (figure \ref{fig:meanvar}a) and the streamwise variance (figure \ref{fig:meanvar}b) for both geometries, which display similar behaviour and magnitudes up to $y^+ \approx 1000$. Beyond that, the $U^+_\phi$-profile in TC flattens out in a bulk flow region which has no correspondence in the channel flow. It is important to note that following \citet{Ostilla2016}, we employ a $z$-dependent mean for TC according to $\langle u_{i}^{\prime 2}\rangle^+ = \langle (u_i^+-\langle u_i^+\rangle_{\phi,t})^2 \rangle_z$, where $\langle \cdot \rangle_i$ denotes an average with respect to $i$. Due to the presence of the Taylor rolls, using the conventional approach of averaging over all directions simultaneously results in a significantly different distribution (cf. the dashed line in figure \ref{fig:meanvar}b). The symbols in figure \ref{fig:meanvar} correspond to the wall-normal positions, $y^+= 30$ and $y^+=90$, for which results for the structure functions will be reported in \S\ref{sec:TCres}.

\subsection{Experimental high-$Re_\tau$ data from TBL and pipe flows}
In order to explore the universality of $E^*_{p,m}$ across Reynolds numbers and flow types, we  employ two sets of high-$Re$ measurements. For the TBL, we resort to the hotwire measurements performed in the Melbourne wind tunnel as first reported in \citet{Hutchins2009}. This dataset encompasses 5 different $Re_\tau$ ranging from 2800 to 19000 at various wall normal locations.
Additionally, we make use of data recorded in the Princeton Superpipe using the so-called NSTAP probe \citep{Vallikivi2011}. These measurements span a range of $2000<Re_{\tau}<98000$ (based on the pipe radius) and are described in more detail in the original publication of   \citet{Hultmark2012}. As both datasets consist of time resolved  point measurements, we make use of Taylor's hypothesis based on the local mean velocity to transform them into the spatial domain. 

\rev{For all datasets employed,} convergence of the structure functions up to tenth order ($p=5$) was verified by plotting the probability density functions $P(\Delta u_i)$ in premultiplied form, i.e. $(\Delta u_i)^pP$,  at various separation distances for all datasets (not shown). The integral of the plotted quantity then represents the desired structure function and its convergence is judged by ensuring that the tails of $(\Delta u_i)^pP$ have converged. This procedure has been employed previously by \citet{Meneveau2013,Huisman2013pre} and others.

\section{Universality of the additive constant $E^*_{p,m}$} \label{sec:ADD}
\begin{figure} 
\psfrag{a}[c][c][1]{$(a)$}
\psfrag{b}[c][c][1]{$(b)$}
\psfrag{c}[c][c][1]{$(c)$}
\psfrag{d}[c][c][1]{$(d)$}
\psfrag{e}[c][c][1]{$(e)$}
\psfrag{f}[c][c][1]{$(f)$}
\psfrag{S2}[c][c][1]{$\SFnew[x]{x}{1}$}
\psfrag{S4}[c][c][1]{$\SFnew[x]{x}{2}$}
\psfrag{S6}[c][c][1]{$\SFnew[x]{x}{3}$}
\psfrag{S10}[c][c][1]{$\SFnew[x]{x}{5}$}
\psfrag{Estar4}[c][c][0.9]{$S_2-D_2/D_1  S_1$}
\psfrag{Estar6}[c][c][1]{$S_3-D_3/D_1S_1$}
\psfrag{Estar10}[c][c][1]{$S_{5}-D_5/D_1S_1$}
\psfrag{zp}[c][c][1]{$y_0^+$}
\centering
\vspace{0.3cm}
      \subfloat{  \begin{tikzpicture}
        \node[anchor=south west,inner sep=0] (image) at (0,0) {\includegraphics[width = \textwidth,trim=0 0 0 0.1cm, clip]{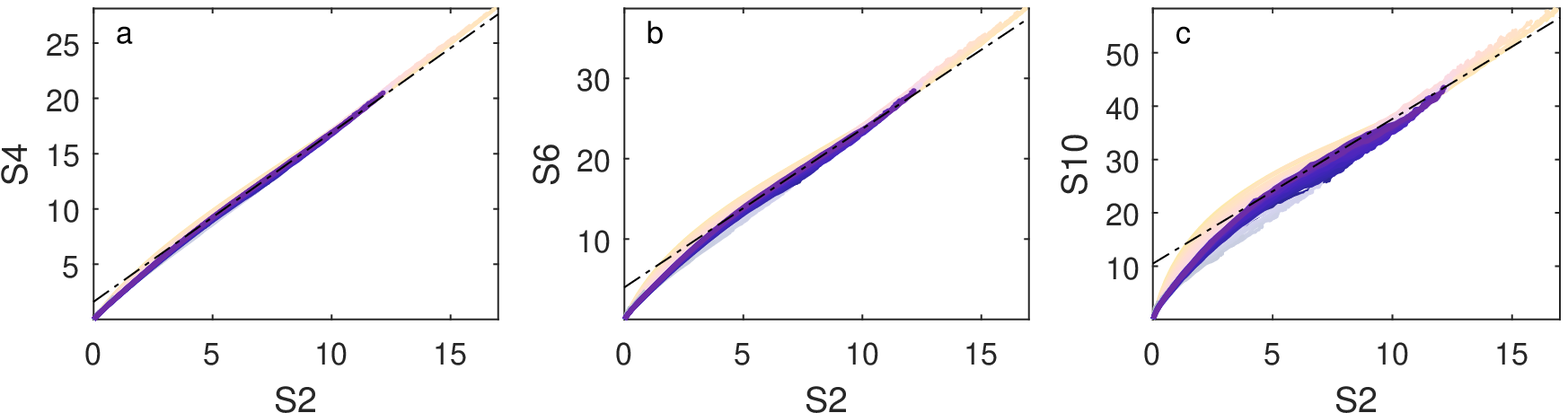}};
        \begin{scope}[x={(image.south east)},y={(image.north west)}]
                               \draw [->] (0.8,0.65) -- (0.83,0.4);
       \node[inner sep=0,anchor=center] (note1) at (0.84,0.38) {\scalebox{1}{$y^+$}};
        \end{scope}
    \end{tikzpicture}}\\ \vspace{0.01cm}
     \subfloat{  \begin{tikzpicture}
        \node[anchor=south west,inner sep=0] (image) at (0,0) {\includegraphics[width = \textwidth]{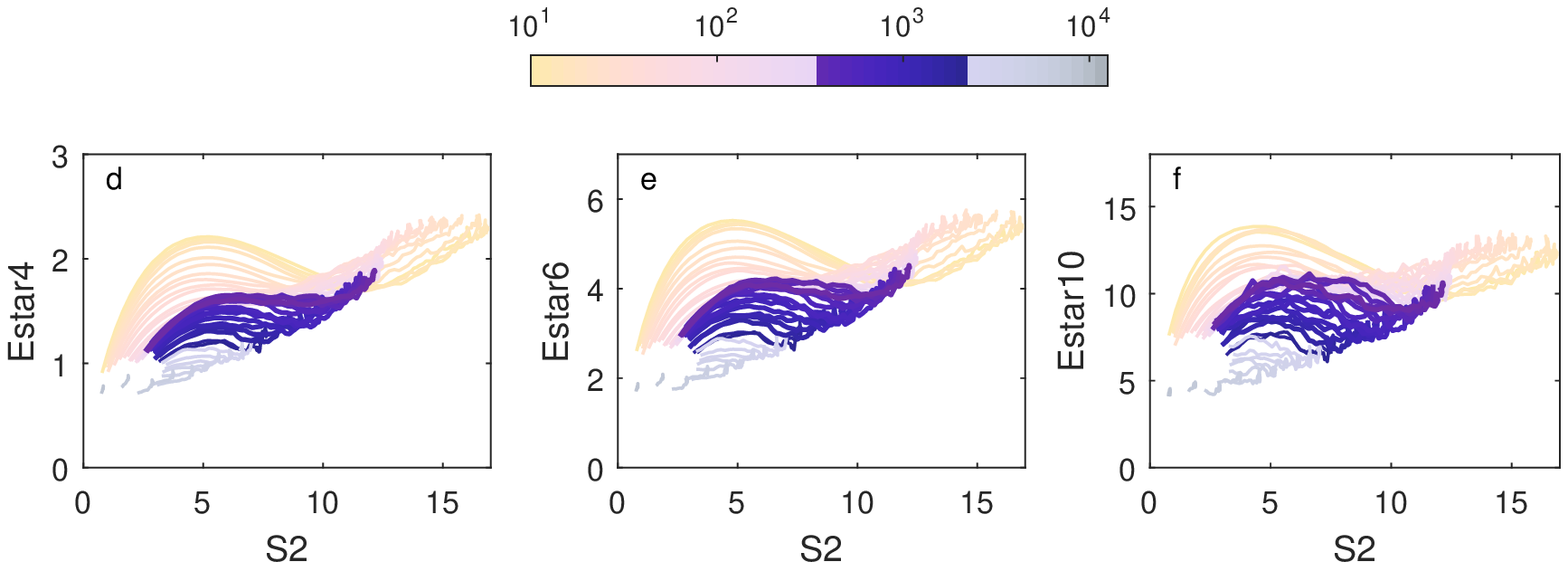}};
        \begin{scope}[x={(image.south east)},y={(image.north west)}]
                \draw[thin,decorate,decoration={brace,amplitude=4pt,mirror}] (0.474+0.05,0.8504) -- (0.570+0.05,0.8504);
                \node[inner sep=0,anchor=center] (note1) at (0.524+0.05,0.7800) {\scalebox{0.86}{log region}};
                \node[inner sep=0,anchor=center] (note1) at (0.315,0.876) {\scalebox{1}{$y_0^+$}};
                       \draw [->] (0.13,0.62) -- (0.13,0.285);
       \node[inner sep=0,anchor=center] (note1) at (0.13,0.24) {\scalebox{1}{$y_0^+$}};

        \end{scope}
    \end{tikzpicture}}\\
    \caption{\label{fig:TBL}(a-c) ESS relations of higher-order structure functions for the range $10<y^+<\delta^+$ computed from the data of \citet{Hutchins2009} at $Re_{\tau}=13700$. Black lines indicate the slopes $D_p/D_1$ reported in \citet{deSilva2017}. (d-e) Compensated form of the ESS relations in (a-c); the range of $\SFnew[x]{x}{1}$ corresponds to  $s_x >y $ for each wall-normal location.}
\end{figure}

In figure \ref{fig:TBL}(a-c), we show longitudinal structure functions  for the TBL at fourth, sixth and tenth order as a function of $\SFnew[x]{x}{1}$ (i.e. we chose $2m=2$ as reference order in (\ref{eq:ESS})). All plots include a range of wall distances from $y^+ =10$ to $y^+=\delta^+=Re_\tau$, spanning almost the entire boundary layer at $Re_\tau = 13700$. The remarkable result of  \citet{deSilva2017} is that  when plotted in this form (instead of vs. separation distance $s_i$), all curves for a given order attain the same slope $D_p/D_m$ beyond $S_1(s_x \approx y_0)$. However, as becomes more evident with increasing order, there is a $y_0^+$-dependence for the additive constants $E^*_{p,1}$. In particular, it is evident for $\langle (\Delta u_x^+)^{10}\rangle^{1/5}$  in figure \ref{fig:TBL}c that $E^*_{5,1}$  decreases with increasing $y_0^+$. In order to further investigate this observation, we plot the  compensated form  \citep[according to (\ref{eq:ESS}) and using the $D_p/D_m$-ratios given by][]{deSilva2017} of the respective structure functions in figures \ref{fig:TBL}(a-c)   in panels (d-f) of the same figure. In this way, deviations from the respective slope $D_p/D_m$ are scrutinized as aberrations from straight horizontal lines. Compared to the absolute values of $\langle (\Delta u_x^+)^{2p}\rangle^{1/p}$, the agreement with (\ref{eq:ESS}) is generally good even outside the logarithmic region \citep[defined using the bounds $3Re_\tau^{1/2}<y^+<0.15Re_\tau$,][]{Marusic2013} with the largest differences occurring  close to the wall. The even more important point in the present context is however that the trend of decreasing $E^*_{p,1}$ with increasing $y_0^+$ emerges for all orders.

\begin{figure} 
\psfrag{a}[c][c][1]{$(a)$}
\psfrag{b}[c][c][1]{$(b)$}
\psfrag{c}[c][c][1]{$(c)$}
\psfrag{d}[c][c][1]{$(d)$}
\psfrag{e}[c][c][1]{$(e)$}
\psfrag{f}[c][c][1]{$(f)$}
\psfrag{S10}[c][c][1]{$S_{10}(u_{x})$}
\psfrag{E2}[c][c][1]{$E^*_{2,1}$}
\psfrag{E3}[c][c][1]{$E^*_{3,1}$}
\psfrag{E5}[c][c][1]{$E^*_{5,1}$}
\psfrag{uvar}[c][c][1]{$\langle u_x^{\prime 2}\rangle^+$}
\psfrag{zp}[c][c][1]{$y_0^+$}
\centering
\vspace{0.5cm}
   {   \subfloat{  \begin{tikzpicture}
        \node[anchor=south west,inner sep=0] (image) at (0,0) {\includegraphics[width = \textwidth]{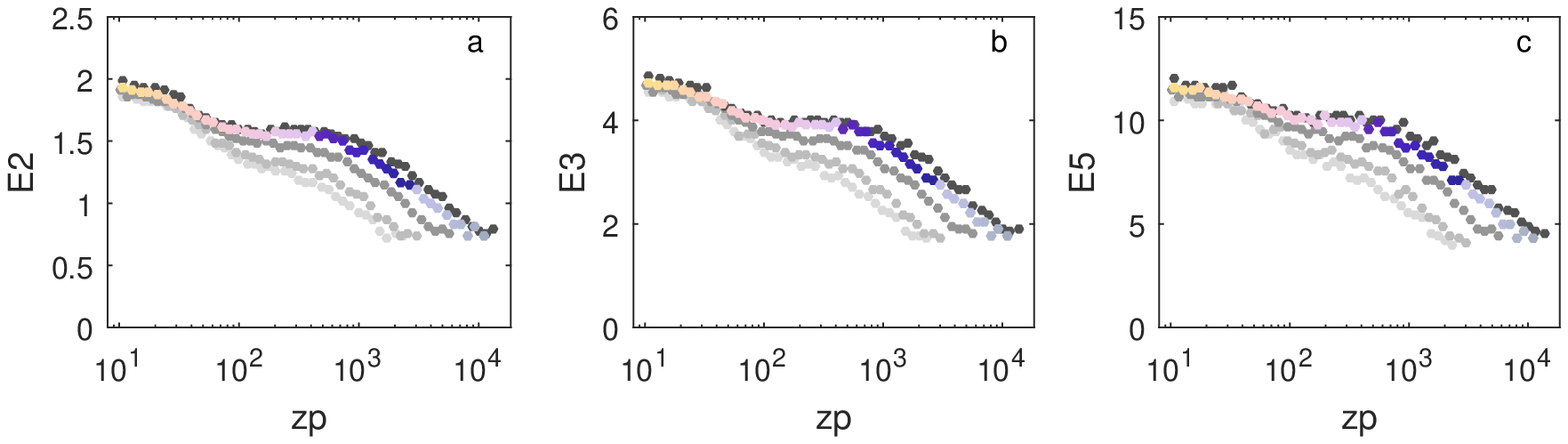}};
        \begin{scope}[x={(image.south east)},y={(image.north west)}]
                \node[inner sep=0,anchor=center] (note1) at (0.2,0.48) {\scalebox{1}{$Re_{\tau}$}};
                \draw [->] (0.23,0.75) -- (0.17,0.48);
        \end{scope}
    \end{tikzpicture}}}\\ 
     {\subfloat{  \begin{tikzpicture}
        \node[anchor=south west,inner sep=0] (image) at (0,0) {\includegraphics[width = \textwidth]{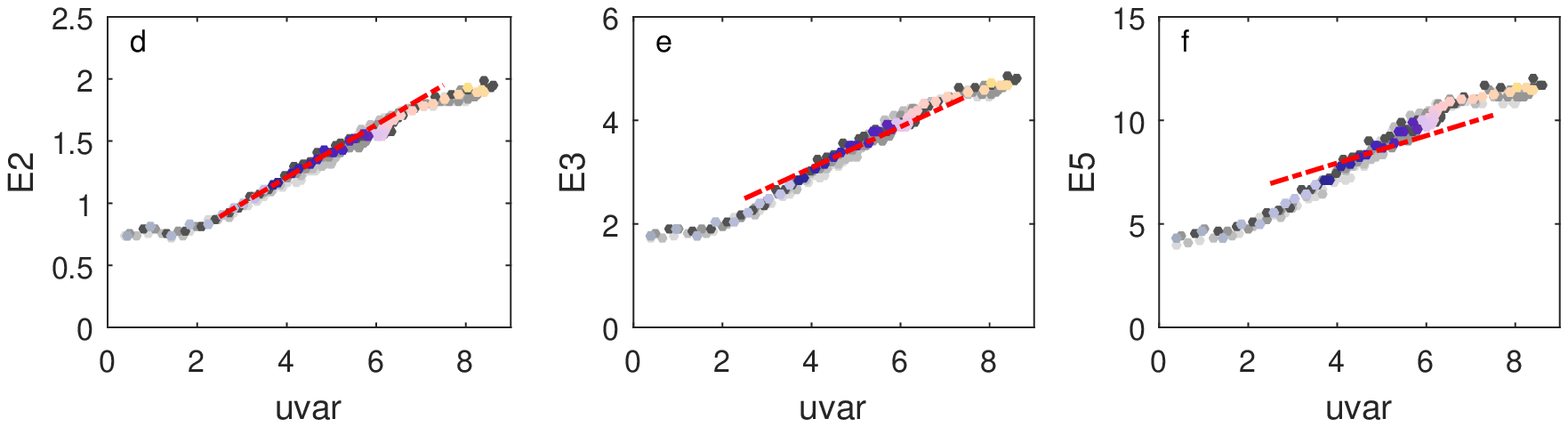}};
    \end{tikzpicture}}}
    \caption{ \label{fig:estar} Additive constant $E^*_{p,m}$ in the ESS form as defined in (\ref{eq:ESS}) as a functions of  $y_0^+$ (a-c) and $\langle u_x^{\prime 2}\rangle^+$ (d-f). In addition to the data at $Re_{\tau} =13700$ (shown using the same colormap as in figure \ref{fig:TBL}), other measurements from the dataset of \citet{Hutchins2009} spanning the range $2800 \leq Re_{\tau} \leq 19000$ are represented by varying shades of grey. Dashed-dotted lines in (d-f) show the slopes suggested by the estimate of (\ref{eq:Estarslope}).   }
\end{figure}

This is also verified in figure \ref{fig:estar}(a-c), where we show $E^*_{p,m}$ taken as the mean of the quantity plotted in figure \ref{fig:TBL}(d-f) as a function of $y_0^+$. Additional data at different Reynolds numbers included in figures \ref{fig:estar}(a-c) reveal that $E^*_{p,1}$ further depends on $Re_\tau$. To the trained eye, this dependency and the shape of the profiles are reminiscent of plots of the streamwise variance and indeed the same data is seen to collapse when plotted vs. the local value of $\langle u_x^{\prime 2}\rangle^+$ in figures \ref{fig:estar}(d-f). Such a behaviour is not purely heuristic, however, as it can be derived by the following considerations: In the limit $s/y\to \infty$ the two points  contributing to $\Delta u$ can be considered independent and $\langle (\Delta u_x^+)^{2p}\rangle^{1/p}\to C_p = \textrm{const.}$ Evaluating (\ref{eq:ESS}) for this case with $m = 1$ yields
\begin{equation}
E^*_{p,1} = C_p -\frac{D_p}{D_1}C_1
\end{equation}
For the second order, we have $C_1 = 2\langle u_x^{\prime 2}\rangle^+$ but generally $C_p$ is a combination of higher order moments of $u_x^{\prime +}$. However, \citet{deSilva2015scaling}
provide the rather simple approximation 
\begin{equation}
C_p \approx G_p \ln\left(\frac{c\delta}{y}\right),
\label{eq:Capprox}
\end{equation}
where $G_p$ are constants given by a combination of the logarithmic slopes of higher order moments of $u_x^{\prime +} $ provided by \citet{Meneveau2013}. 
Making use of this approximation and employing the  log-law for $
\langle u_x^{\prime 2}\rangle^+ = B_1-A_1 \ln \left( \frac{y}{\delta}\right)
$
we arrive at 
\begin{equation}
E^*_{p,1} =  \left(\frac{G_p}{A_1}- 2\frac{D_p}{D_1} \right)\langle u_x^{\prime 2}\rangle^+ +\textrm{const.}
\label{eq:Estarslope}
\end{equation}
We note that the choice $m =1$ is not a prerequisite for this form and a similar expression can be derived for arbitrary references $m$.
The slopes suggested by the linear relation (\ref{eq:Estarslope}) are included as red dashed-dotted lines in figure \ref{fig:estar}(d-f). While the agreement with the data is excellent for $p=1$ even slightly beyond the logarithmic region, for which (\ref{eq:Estarslope}) was derived, conformance becomes  somewhat worse with increasing order. This is consistent with the fact that  the approximation (\ref{eq:Capprox}) also becomes progressively weaker with increasing $p$ \citep[cf.][]{deSilva2015scaling}. Nevertheless, the collapse of $E^*_{p,1}$ on $\langle u_x^{\prime 2}\rangle^+$ is robust for all orders and across the entire boundary layer. We will revisit this result and examine universality across flow geometries in \S\ref{sec:Estar_geom}.

\section{Comparison of higher-order structure function in Taylor-Couette and channel flow} \label{sec:TCres}

\subsection{Streamwise velocity component}

\begin{figure} 
\psfrag{a}[c][c][1]{$(a)$}
\psfrag{b}[c][c][1]{$(b)$}
\psfrag{c}[c][c][1]{$(c)$}
\psfrag{d}[c][c][1]{$(d)$}
\psfrag{e}[c][c][1]{$(e)$}
\psfrag{f}[c][c][1]{$(f)$}
\psfrag{sony}[c][c][1]{$s_{\phi,x}/y_0$}
\psfrag{stony}[c][c][1]{$s_{z}/y_0$}
\psfrag{S2}[c][c][1]{$\SFnew[x]{x,\phi}{1}$}
\psfrag{S4}[c][c][1]{$\SFnew[x]{x,\phi}{2}$}
\psfrag{S10}[c][c][1]{$\SFnew[x]{x,\phi}{5}$}
\psfrag{S2T}[c][c][1]{$\SFTnew[z]{x,\phi}{1}$}
\psfrag{S4T}[c][c][1]{$\SFTnew[z]{x,\phi}{2}$}
\psfrag{S10T}[c][c][1]{$\SFTnew[z]{x,\phi}{5}$}
\psfrag{lam}[c][c][0.8]{$\lambda_{TR}/(2y_0)$}
\psfrag{CH}[l][l][0.8]{CH}
\psfrag{TC}[l][l][0.8]{TC}
\psfrag{yp1}[l][l][0.8]{$y_0^+=30$}
\psfrag{yp2}[l][l][0.8]{$y_0^+=90$}
    \centering
    \vspace{0.3cm}
   \subfloat{ \begin{tikzpicture}
        \node[anchor=south west,inner sep=0] (image) at (0,0) {\includegraphics[scale = 0.75]{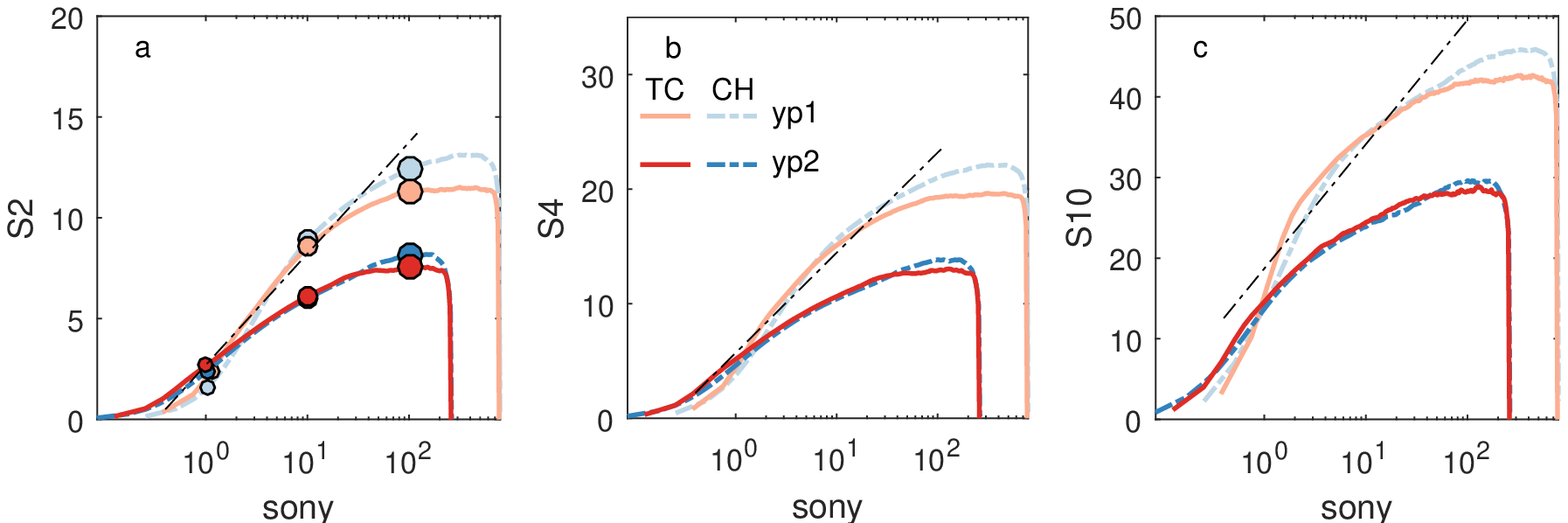}};
        \begin{scope}[x={(image.south east)},y={(image.north west)}]
            \node[anchor=south west,inner sep=0] (image) at (0.075,0.6) {\includegraphics[scale= 0.32]{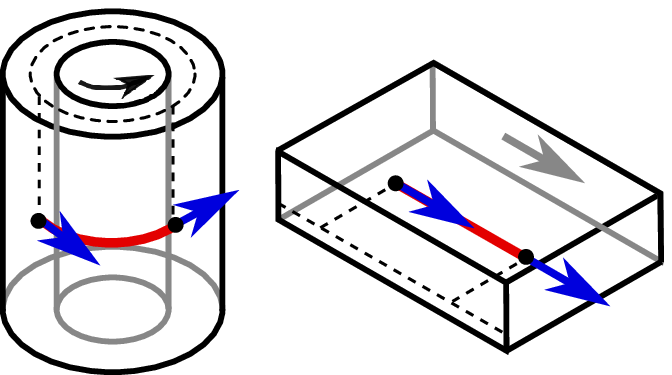}};
        \end{scope}
    \end{tikzpicture}}\\\vspace{0.3cm}
       \subfloat{ \begin{tikzpicture}
        \node[anchor=south west,inner sep=0] (image) at (0,0) {\includegraphics[scale = 0.75]{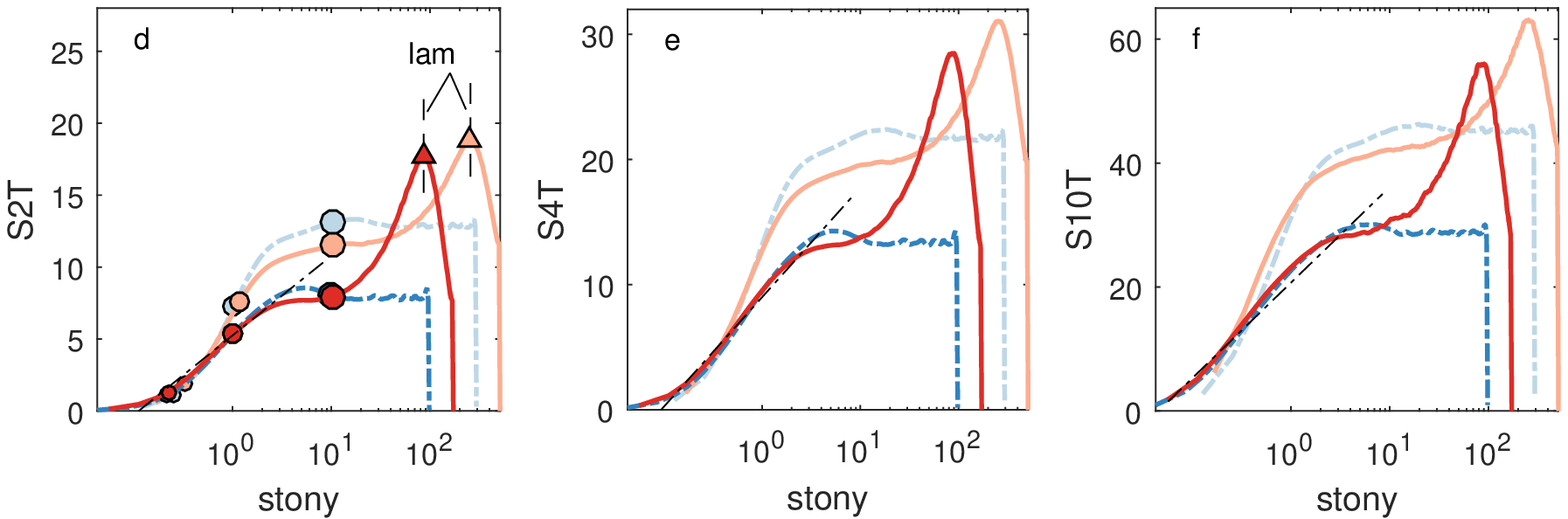}};
        \begin{scope}[x={(image.south east)},y={(image.north west)}]
            \node[anchor=south west,inner sep=0] (image) at (0.075,0.6) {\includegraphics[scale= 0.32]{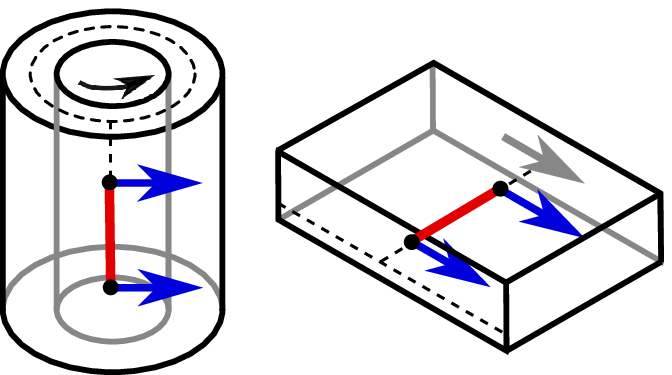}};
        \end{scope}
    \end{tikzpicture}}\\
    \caption{\label{fig:udir} Longitudinal (a-c) and transversal (d-f) structure function of the streamwise (azimutal) velocity component $u_x$ ($u_{\phi}$) for Taylor Couette (shades of red) and Channel (shades of blue). Two different wall-normal positions are shown for structure functions of second (a,d), fourth (b,d) and 10th order (c,f). Pictograms in (a,d) illustrate the geometry of the respective structure function and the black lines represent  logarithmic slopes at the  respective order for the longitudinal direction from \citet{deSilva2015scaling}. The legend in (b) applies to all panels. }
\end{figure}

We start out by considering structure functions of the streamwise velocity components in the conventional form as a function of $s_i/y_0$. Results for the longitudinal direction, which is the most accessible experimentally, are plotted in figures \ref{fig:udir}(a-c) at increasing orders. For both TC and channel flow data at $y_0^+ = 30$ and $y_0^+ = 90$ are displayed. After an initial increase, all curves are observed to level off at a constant value. The subsequent drop back to zero is an artefact of the periodic boundary conditions used in the simulations, which implies periodicity also for the structure functions. \citet{deSilva2015scaling} were able to determine the logarithmic slopes $D_p$ in (\ref{eq:logSF}) directly based on high-$Re_\tau$ data. Comparing their results (included as black lines in figures \ref{fig:udir}a-c) to the present data clearly shows that the scaling according to (\ref{eq:logSF}) does not hold for either of the data sets or wall-normal positions plotted. Apart from not following a straight line, the structure functions also exhibit distinctively different slopes at different wall-normal positions $y_0$. The differences between TC and channel flow in figures \ref{fig:udir}a-c are minimal conforming with the good agreement observed for the $\var[x]$-profile in figure \ref{fig:meanvar}b.

Similar observations can be made for the transversal structure functions of $u_{\phi,x}$ displayed at increasing orders in figures \ref{fig:udir}(d-f). Generally, the initial  increase is steeper for $\SFTnew[z]{\phi,x}{p}$ and the curves level off at lower normalized distances $s_z/y_0$. In the attached eddy picture, this implies that the eddies have an aspect ratio $R_x/R_z>1$. Using the slopes $D_p$ determined by \citet{deSilva2015scaling} for the longitudinal component (black lines in figures \ref{fig:udir}(d-f)) as reference, it becomes clear that indeed the increase of $\SFTnew[z]{\phi,x}{p}$ is steeper than that of $\SFnew[x]{x}{p}$ at the same $y_0$. This observation is in line with the predictions in \S\ref{sec:AE} and with findings in the spectral domain by  \citet{Lee2015} and \citet{Chandran2017}. However, the striking feature of the transversal structure functions is a pronounced peak at large separations that is uniquely observed in TC flow. This peak occurs at values of $s_z$ for which turbulent fluctuations have already become uncorrelated (i.e. the curves have levelled off). Based on this along with the fact that the locations of the peaks correspond to $\lambda_{TR}$ as indicated in figure \ref{fig:udir}d, the peaks can be related to the existence of high and low speed streaks induced by the Taylor rolls. These velocity differences are substantial. From the fact that the maximum of $\SFnew[x]{x}{1}$ is located at about twice the turbulence level, their magnitude can be estimated to be $\approx (2\var[\phi])^{1/2}$ at the investigated wall-normal positions. Also here, the eventual decline to zero is an artefact of the finite box size; further peaks above the turbulence level with a spacing of $\lambda_{TR}$ would be expected for domains containing additional pairs of rollers.

\begin{figure} 
\psfrag{a}[c][c][1]{$(a)$}
\psfrag{b}[c][c][1]{$(b)$}
\psfrag{c}[c][c][1]{$(c)$}
\psfrag{d}[c][c][1]{$(d)$}
\psfrag{e}[c][c][1]{$(e)$}
\psfrag{f}[c][c][1]{$(f)$}
\psfrag{sony}[c][c][1]{$s_{\phi}/y_0,\, s_{x}/y_0$}
\psfrag{stony}[c][c][1]{$s_{z}/y_0$}
\psfrag{S2}[c][c][1]{$\SFnew[x]{x,\phi}{1}$}
\psfrag{S4}[c][c][1]{$\SFnew[x]{x,\phi}{2}$}
\psfrag{S6}[c][c][1]{$\SFnew[x]{x,\phi}{3}$}
\psfrag{S10}[c][c][1]{$\SFnew[x]{x,\phi}{5}$}
\psfrag{S2T}[c][c][1]{$\SFTnew[z]{x,\phi}{1}$}
\psfrag{S4T}[c][c][1]{$\SFTnew[z]{x,\phi}{2}$}
\psfrag{S6T}[c][c][1]{$\SFTnew[z]{x,\phi}{3}$}
\psfrag{S10T}[c][c][1]{$\SFTnew[z]{x,\phi}{5}$}
\psfrag{CH}[l][l][0.8]{CH}
\psfrag{TC}[l][l][0.8]{TC}
\psfrag{yp1}[l][l][0.8]{$y_0^+=30$}
\psfrag{yp2}[l][l][0.8]{$y_0^+=90$}
    \centering
    \vspace{0.3cm}
   \subfloat{ \begin{tikzpicture}
        \node[anchor=south west,inner sep=0] (image) at (0,0) {\includegraphics[scale = 0.75]{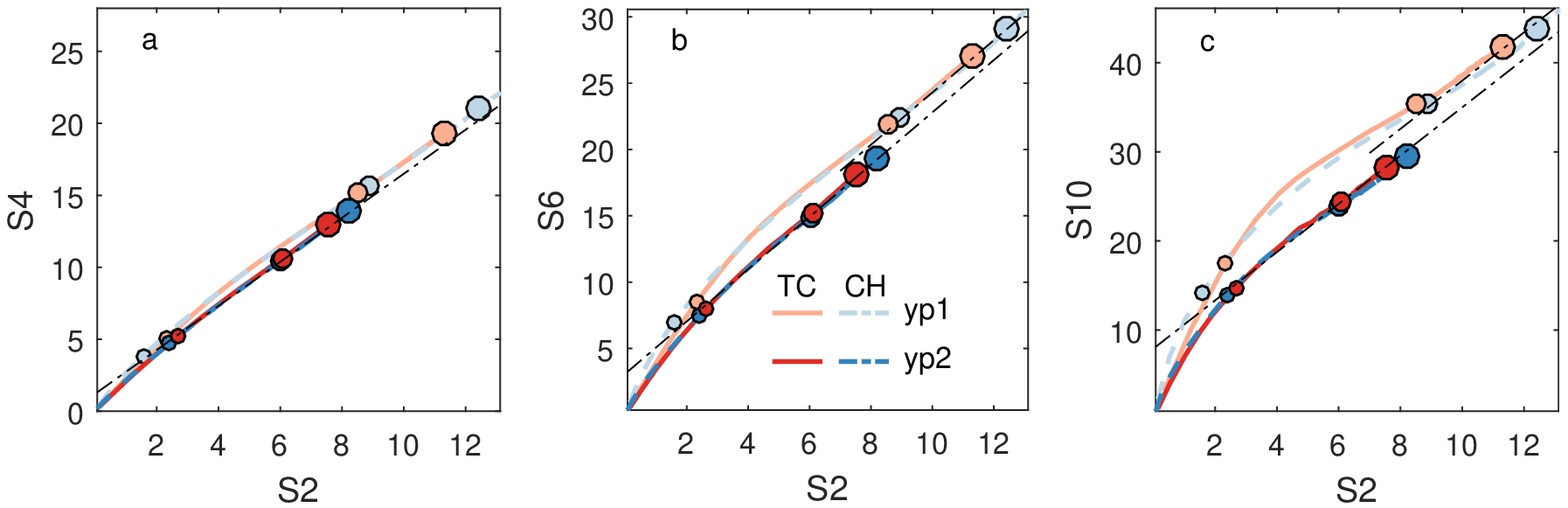}};
        \begin{scope}[x={(image.south east)},y={(image.north west)}]
            \node[anchor=south west,inner sep=0] (image) at (0.075,0.6) {\includegraphics[scale= 0.32]{figures_TCsfun/Fig6a_7a_inset.eps}};
        \end{scope}
    \end{tikzpicture}}\\ \vspace{0.2cm}
       \subfloat{ \begin{tikzpicture}
        \node[anchor=south west,inner sep=0] (image) at (0,0) {\includegraphics[scale = 0.75]{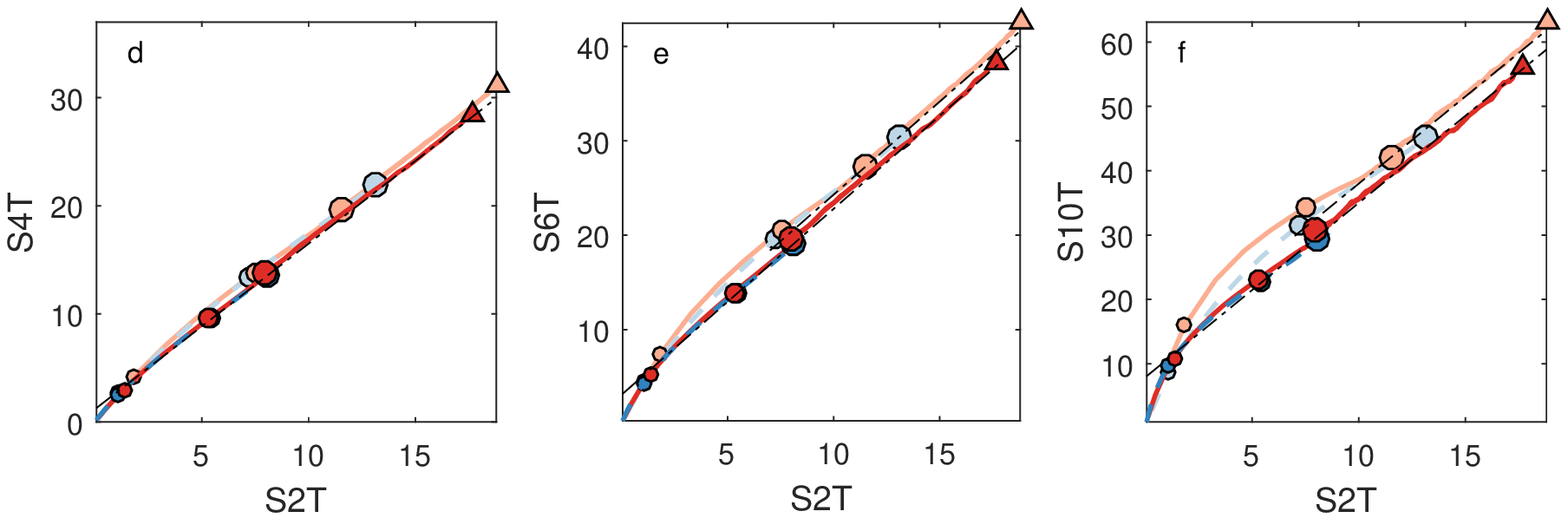}};
        \begin{scope}[x={(image.south east)},y={(image.north west)}]
            \node[anchor=south west,inner sep=0] (image) at (0.075,0.6) {\includegraphics[scale= 0.32]{figures_TCsfun/Fig6d_7d_inset.eps}};
        \end{scope}
    \end{tikzpicture}}\\
    \caption{\label{fig:uESS} Higher-order structure functions plotted vs. second order for the streamwise/azimuthal velocity component in longitudinal (a-c) and transversal direction (d-f). Circles of increasing size mark locations where $s_{\phi,x}/y_0$ is equal to 1, 10 and 100 (a-c) or where $s_z/y_0$ is 0.2, 1 and 10 (d-e). The triangles in (d-f) mark the position of the Taylor-roll peak. All these positions are also marked in figure \ref{fig:udir}(a,c) for reference. Black lines indicate the respective values of $D_p/D_m$ determined by \citet{deSilva2017} for the longitudinal structure function of $u_x$ in TBL flow.    }
\end{figure}

Next, we return to the analysis in the spirit of the ESS method. For this purpose, we present higher order structure functions in the longitudinal direction with reference to $\SFnew[x]{x}{1}$ 
(in figures \ref{fig:uESS}(a-c)) and with reference to $\SFTnew[z]{x}{1}$ (in figure \ref{fig:uESS}(d-f)), which were displayed in figures \ref{fig:udir}(a),(d), respectively. In order  to establish the correspondence to the separation distance, we mark values of $\SFnew[x]{x}{1}$ and $\SFTnew[z]{x}{1}$ corresponding to $s_{\phi,x}/y_0 = 1,10,100$ and $s_{z}/y_0 = 0.2,1,10$, respectively, with circles of increasing size in figure \ref{fig:uESS} which are also shown in figures \ref{fig:udir}(a),(d). Additionally, we mark the location of the peak in figures \ref{fig:udir}(d-f) with a triangle. Altogether, good agreement with the slopes measured in \citet{deSilva2017} (indicated by black lines) is observed for the longitudinal direction in figure \ref{fig:uESS}(a-b) at $s_{\phi,x}/y_0>1$ for both channel and TC. The only small limitation is at the position closer to the wall ($y_0^+=30$), where due to an initial `overshoot' whose intensity increases with order $p$, the slope is only recovered at slightly higher values of $r/y_0$. A similar effect is also present in the TBL data as evidenced by the small `hump' at low values of $S_1$ for small $y_0^+$  in figure \ref{fig:TBL}.

Consistent with the above discussion on the aspect ratio of attached eddies, the plots for the transversal direction in figures \ref{fig:uESS}(d-f) attain the respective slopes $D_p/D_1$ at lower multiples of $y_0$ in both flows considered. Besides that, it is remarkable in view of the distinctively different behaviour in figures \ref{fig:udir}(d-f) how well the Taylor-roll peaks align with those slopes even for the tenth order moment in figure \ref{fig:uESS}(d).

\subsection{Spanwise velocity component}
\begin{figure} 
\psfrag{a}[c][c][1]{$(a)$}
\psfrag{b}[c][c][1]{$(b)$}
\psfrag{c}[c][c][1]{$(c)$}
\psfrag{d}[c][c][1]{$(d)$}
\psfrag{e}[c][c][1]{$(e)$}
\psfrag{f}[c][c][1]{$(f)$}
\psfrag{sony}[c][c][1]{$s_{\phi,x}/y_0$}
\psfrag{stony}[c][c][1]{$s_{z}/y_0$}
\psfrag{S2}[c][c][1]{$\SFnew[z]{z}{1}$}
\psfrag{S4}[c][c][1]{$\SFnew[z]{z}{2}$}
\psfrag{S10}[c][c][1]{$\SFnew[z]{z}{5}$}
\psfrag{S2T}[c][c][1]{$\SFTnew[x]{z}{1}$}
\psfrag{S4T}[c][c][1]{$\SFTnew[x]{z}{2}$}
\psfrag{S10T}[c][c][1]{$\SFTnew[x]{z}{5}$}
\psfrag{CH}[l][l][0.8]{CH}
\psfrag{TC}[l][l][0.8]{TC}
\psfrag{yp1}[l][l][0.8]{$y^+=30$}
\psfrag{yp2}[l][l][0.8]{$y^+=90$}
\psfrag{lam}[c][c][0.8]{$\lambda_{TR}/(2y_0)$}
    \centering
    \vspace{0.3cm}
   \subfloat{ \begin{tikzpicture}
        \node[anchor=south west,inner sep=0] (image) at (0,0) {\includegraphics[scale = 0.75]{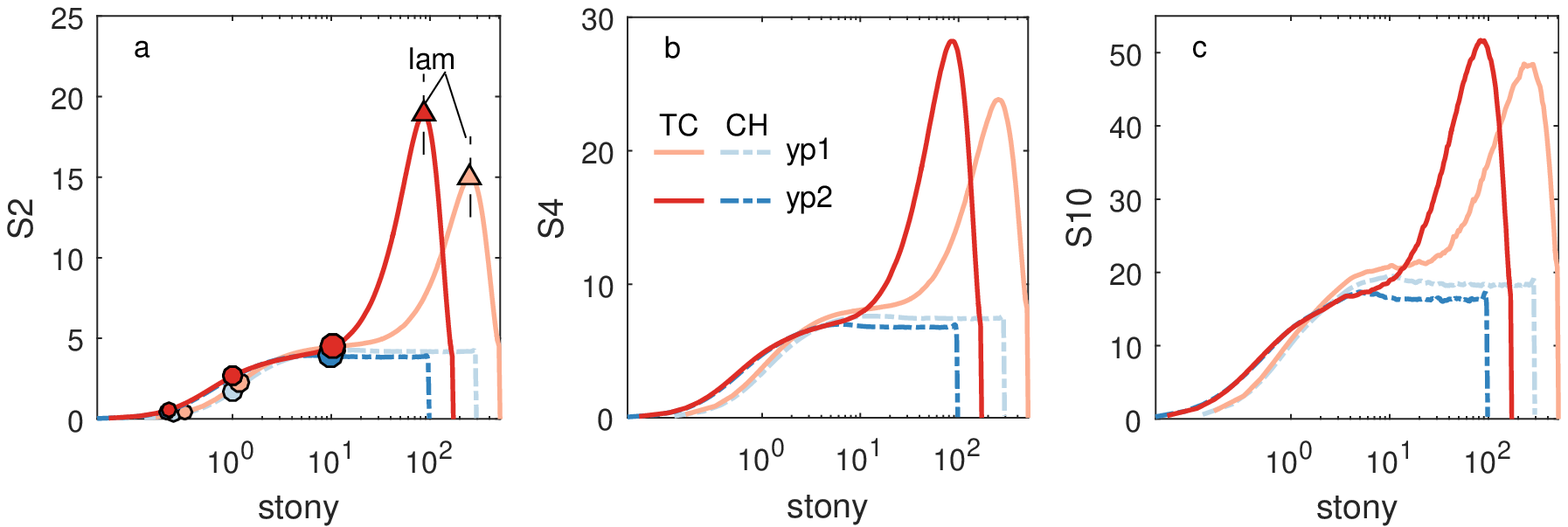}};
        \begin{scope}[x={(image.south east)},y={(image.north west)}]
            \node[anchor=south west,inner sep=0] (image) at (0.075,0.6) {\includegraphics[scale= 0.32]{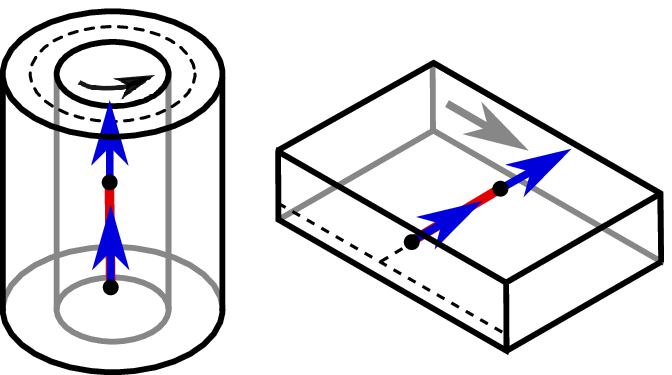}};
        \end{scope}
    \end{tikzpicture}}\\ 
       \subfloat{ \begin{tikzpicture}
        \node[anchor=south west,inner sep=0] (image) at (0,0) {\includegraphics[scale = 0.75]{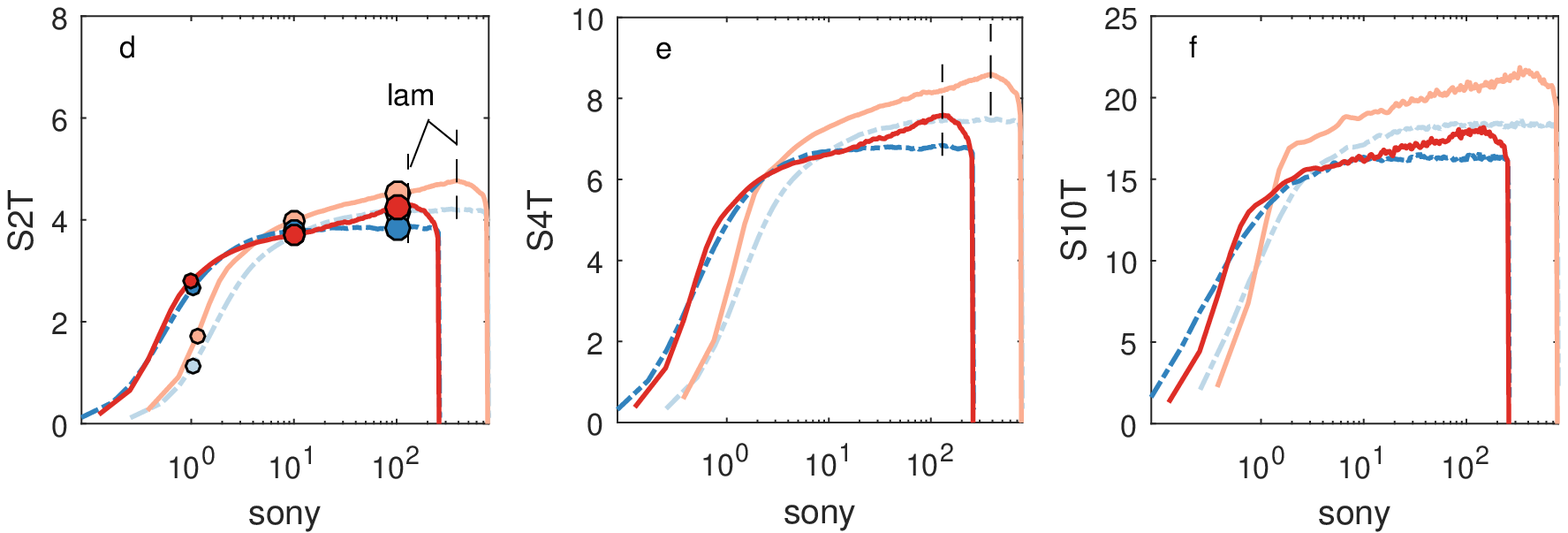}};
        \begin{scope}[x={(image.south east)},y={(image.north west)}]
            \node[anchor=south west,inner sep=0] (image) at (0.075,0.6) {\includegraphics[scale= 0.32]{figures_TCsfun/Fig8a_9a_inset.eps}};
        \end{scope}
    \end{tikzpicture}}\\
    \caption{\label{fig:vdir} Same as figure \ref{fig:udir} for the spanwise velocity component $u_z$. The legend in (b) applies to all panels. Circles of increasing size mark locations where $s_z/y_0$ is 0.2, 1 and 10  (a-c) or where $s_{\phi,x}/y_0$ is equal to 1, 10 and 100 (d-e). }
\end{figure}

While structure functions in both directions of the streamwise component were already considered in \citet{deSilva2017} for TBLs, we go beyond their work by investigating the spanwise velocity component in the following. Results for the longitudinal structure functions $\SFnew[z]{z}{p}$, where the separation distance is along the spanwise direction of the flow, are plotted in figures \ref{fig:vdir}(a-c). At a first glance, these plots look very similar to the ones for $u_{\phi,x}$ in the transversal directions (figures \ref{fig:udir}(d-f)) with close agreement between channel and TC before once again the Taylor rolls lead to a sharp peak at $s_z = \lambda_{TR}/2$. The important difference between the peaks in $\SFTnew[z]{\phi}{p}$ and $\SFnew[z]{z}{p}$ is that in the latter case, their occurrence peak is a direct effect of the Taylor rolls, i.e. the rolls contribute directly to $u_z$. This  in contrast to the previous case where their action was indirect in transporting low velocity fluid away from the wall and vice-versa. A noticeable consequence is that the relative magnitude of the peaks is now inverted such that the peak closer to the wall, where the contribution of the Taylor roll is apparently reduced, is lower for $\SFnew[z]{z}{p}$.

The transversal direction for $u_z$ runs along the axis of the Taylor rolls. Hence, their mean contribution cancels out and the structure functions $\SFTnew[x]{z}{p}$ in figures \ref{fig:vdir}(d-f) exhibit no distinct peaks for TC flow. It is however noteworthy that the curves for $\SFTnew[x]{z}{p}$ in TC keep increasing even at distances at which their counterparts in channel flow have already saturated. This can be explained by the fact that the Taylor rolls fluctuate in size and position \citep{Ostilla2015} which introduces correlations at scales larger than those of the turbulence field. The longest correlation possible is limited by half the box size; thus the fact that the end of the increase of $\SFTnew[x]{z}{p}$ coincides with $\lambda_{TR}$ is simply a consequence of choosing the domain size equal to $\lambda_{TR}$. It should therefore not be given physical significance beyond demonstrating that modulations of the Taylor rolls with wavelengths up to at least this size are possible.

\begin{figure} 
\psfrag{a}[c][c][1]{$(a)$}
\psfrag{b}[c][c][1]{$(b)$}
\psfrag{c}[c][c][1]{$(c)$}
\psfrag{d}[c][c][1]{$(d)$}
\psfrag{e}[c][c][1]{$(e)$}
\psfrag{f}[c][c][1]{$(f)$}
\psfrag{sony}[c][c][1]{$s_{\phi}/y_0,\, s_{x}/y_0$}
\psfrag{stony}[c][c][1]{$s_{z}/y_0$}
\psfrag{S2}[c][c][1]{$\SFnew[z]{z}{1}$}
\psfrag{S4}[c][c][1]{$\SFnew[z]{z}{2}$}
\psfrag{S6}[c][c][1]{$\SFnew[z]{z}{3}$}
\psfrag{S10}[c][c][1]{$\SFnew[z]{z}{5}$}
\psfrag{S2T}[c][c][1]{$\SFTnew[x]{z}{1}$}
\psfrag{S4T}[c][c][1]{$\SFTnew[x]{z}{2}$}
\psfrag{S6T}[c][c][1]{$\SFTnew[x]{z}{3}$}
\psfrag{S10T}[c][c][1]{$\SFTnew[x]{z}{5}$}
\psfrag{CH}[l][l][0.8]{CH}
\psfrag{TC}[l][l][0.8]{TC}
\psfrag{yp1}[l][l][0.8]{$y^+=30$}
\psfrag{yp2}[l][l][0.8]{$y^+=90$}
    \centering
    \vspace{0.3cm}
   \subfloat{ \begin{tikzpicture}
        \node[anchor=south west,inner sep=0] (image) at (0,0) {\includegraphics[scale = 0.75]{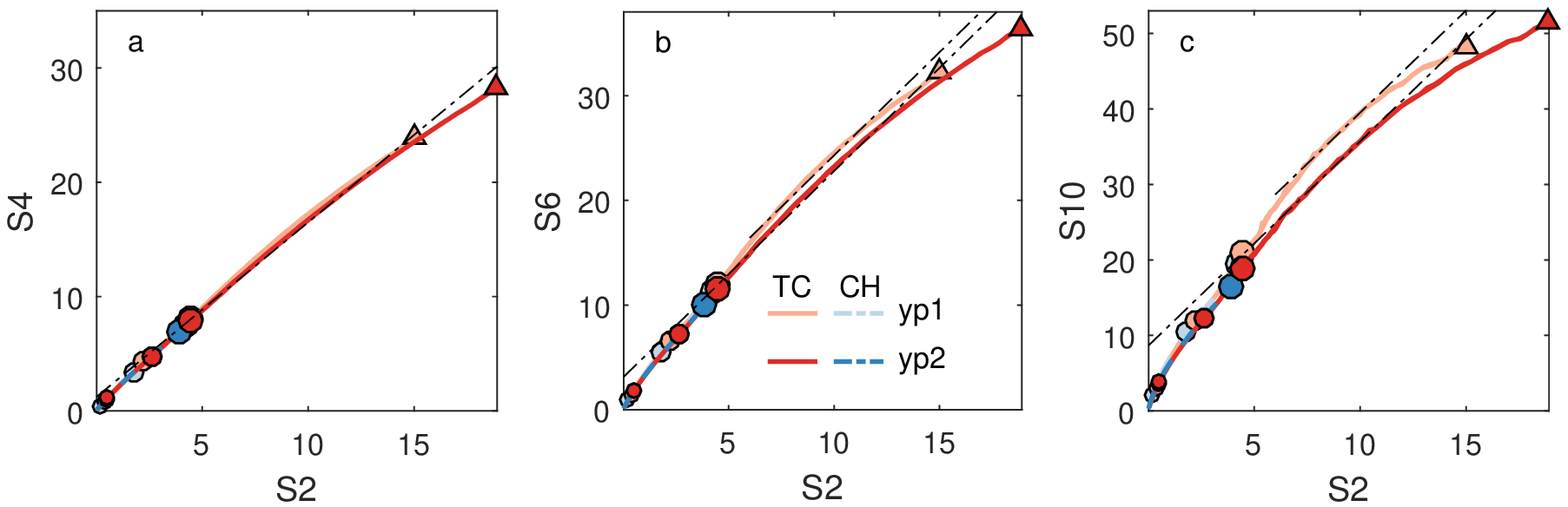}};
        \begin{scope}[x={(image.south east)},y={(image.north west)}]
            \node[anchor=south west,inner sep=0] (image) at (0.075,0.6) {\includegraphics[scale= 0.32]{figures_TCsfun/Fig8a_9a_inset.eps}};
        \end{scope}
    \end{tikzpicture}}\\ 
       \subfloat{ \begin{tikzpicture}
        \node[anchor=south west,inner sep=0] (image) at (0,0) {\includegraphics[scale = 0.75]{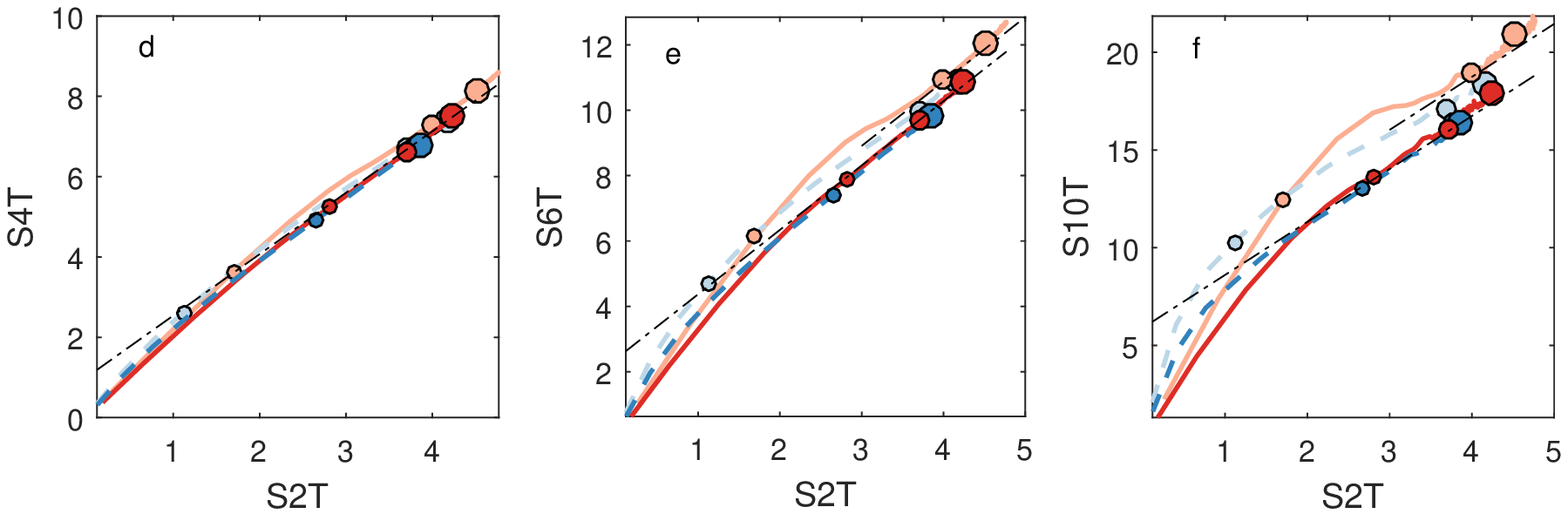}};
        \begin{scope}[x={(image.south east)},y={(image.north west)}]
            \node[anchor=south west,inner sep=0] (image) at (0.075,0.6) {\includegraphics[scale= 0.32]{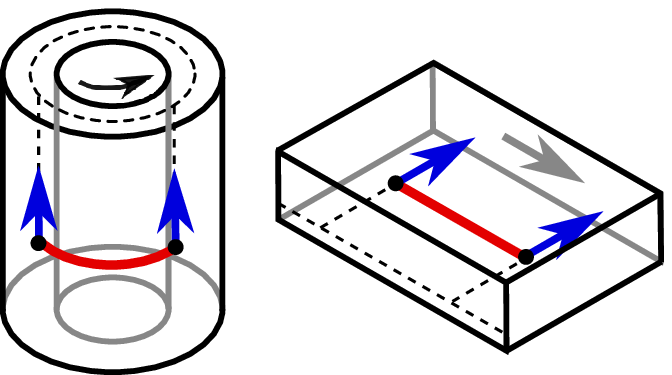}};
        \end{scope}
    \end{tikzpicture}}\\
    \caption{\label{fig:vESS} Same as figure \ref{fig:uESS} for the spanwise velocity component $u_z$. The legend in (b) applies to all panels.}
\end{figure}

The plots of figure \ref{fig:vdir} for the spanwise velocity component are repeated in figure \ref{fig:vESS} in ESS-form. For the longitudinal direction (figures \ref{fig:vESS}a-c), it remains inconclusive whether or not the channel results attain a linear relationship due to the limited scaling range. In the case of TC flow, however, it is obvious that the description by (\ref{eq:ESS}) does not hold since the slope of the curves at both wall-normal positions  is continuously changing. It appears likely that this breakdown is related to the fact that for $\SFnew[z]{z}{p}$ the direct contribution of the Taylor rolls does not cancel out. Consequently, the statistics are influenced by structures that are not directly related to the presence of the wall which may be responsible for deviation from the ECR-scaling observed in other cases.

In agreement with such a reasoning, it is evident from figures \ref{fig:vESS}(d-f) that $\SFTnew[x]{z}{p}$, for which the $u_z$ contribution of the Taylor rolls does cancel out largely, complies with the linear scaling of (\ref{eq:ESS}) for both channel and TC. Even more so, the slopes in the data for the spanwise velocity component seem consistent with those measured for $u_x$ by \cite{deSilva2017}. Thus, there is evidence that at least in the transversal direction, also the structure function of the spanwise velocity adheres to the same universality as $u_{\phi,x}$.

\subsection{Universality of the ratios $D_p/D_1$} \label{sec:TCresD}
\begin{figure} 
\psfrag{a}[c][c][1]{$(a)$}
\psfrag{b}[c][c][1]{$(b)$}
\psfrag{Drel}[c][c][1]{$D_p/D_1$}
\psfrag{p}[c][c][1]{$p$}
\psfrag{CH}[c][c][1]{CH}
\psfrag{TC}[c][c][1]{TC}
\psfrag{TBL}[c][l][1]{TBL}
\psfrag{long}[l][l][1]{long.}
\psfrag{trans}[l][l][1]{trans.}
\psfrag{stream}[l][l][1]{$u_{\phi,x}$ (streamwise)}
\psfrag{spanwise}[l][l][1]{$u_z$ (spanwise)}
\psfrag{eq}[l][l][1]{Eq. (\ref{eq:du2du-ESS})}
\centering
\vspace{0.2cm}
{\includegraphics[scale = 0.75]{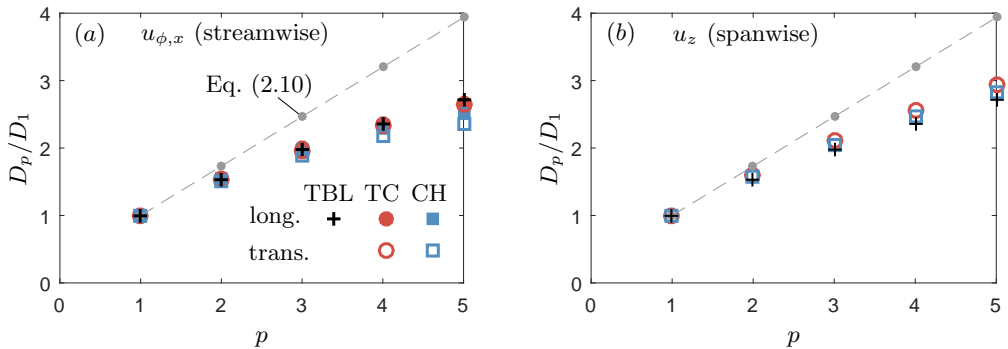}}
\caption{\label{fig:Drel} Result from fitting the slopes $D_p/D_1$ in the ESS form for the streamwise (a) and spanwise (b) velocity component in the longitudinal and transversal direction. Fits are computed for a range corresponding to $s_i/y>1$ at $y_0^+=90$. The legend in (a) applies to both panels, however, only the transversal direction is shown in (b). The boundary layer result is taken from \citet{deSilva2017}; grey dots represent the Gaussian prediction from (\ref{eq:du2du-ESS}).}
\end{figure}

In order to address the question of universality for the slopes in the ESS-form in a more structured manner, we plot results for $D_p/D_1$ in figure \ref{fig:Drel}. In all cases, the data were fitted over a range corresponding to $s_i/y_0>1$ and at $y_0^+=90$ where the scaling is generally  more robust. The TBL results of \citet{deSilva2017} included here are obtained at $Re_\tau =19000$ and therefore also serve as a high-$Re_\tau$ reference for the present numerical datasets. The most obvious result in this plot is that the actual slopes in all geometries considered significantly deviate from the Gaussian estimates of (\ref{eq:du2du-ESS}) for both the streamwise (figure \ref{fig:Drel}a) and the spanwise (b) velocity component. This is similar to \citet{Meneveau2013} who observed sub-Gaussian behaviour for higher order moments of the velocity fluctuations $u^\prime$. Our results indicate that also the distributions of $\Delta u_i$ and $\Delta^T u_i$ are sub-Gaussian.
 Figure \ref{fig:Drel}a reinforces the result from figure \ref{fig:uESS} that for the streamwise velocity the ratios $D_p/D_1$ agree closely up to tenth order for longitudinal as well as transversal separations. This strongly supports universality of this characteristic across all geometries considered --- now also including TC.

For the spanwise velocity component $u_z$, the scaling for the longitudinal direction remained either inconclusive (channel) or - in violation of (\ref{eq:ESS}) - a clearly nonlinear relation was found (TC). Therefore, only the results for $\SFTnew[x]{z}{p}$ are displayed in figure \ref{fig:Drel}b. For these results, the agreement with the TBL results for $u_x$ is good for both channel and TC flow and it may be concluded that the universality also extends to the spanwise velocity component in at least the transversal direction.

\subsection{Universality of $E^*_{p,m}$ across flow geometries}\label{sec:Estar_geom}
\begin{figure} 
\psfrag{a}[c][c][1]{$(a)$}
\psfrag{b}[c][c][1]{$(b)$}
\psfrag{c}[c][c][1]{$(c)$}
\psfrag{d}[c][c][1]{$(d)$}
\psfrag{e}[c][c][1]{$(e)$}
\psfrag{f}[c][c][1]{$(f)$}
\psfrag{S2}[c][c][1]{$S_2(u_{x})$}
\psfrag{S4}[c][c][1]{$S_4(u_{x})$}
\psfrag{S6}[c][c][1]{$S_6(u_{x})$}
\psfrag{S10}[c][c][1]{$S_{10}(u_{x})$}
\psfrag{E2}[c][c][1]{$E^*_{2,1}$}
\psfrag{E3}[c][c][1]{$E^*_{3,1}$}
\psfrag{E5}[c][c][1]{$E^*_{5,1}$}
\psfrag{uvar}[c][c][1]{$\langle u_x^{\prime 2}\rangle^+$}
\psfrag{zp}[c][c][1]{$z^+$}
\psfrag{pipe}[l][l][0.8]{Superpipe}
\psfrag{TC}[l][l][0.8]{TC}
\psfrag{CH}[l][l][0.8]{CH}
\psfrag{long}[l][l][0.8]{$u_{\phi,x}\,$long.}
\psfrag{trans}[l][l][0.8]{$u_{\phi,x}\,$trans.}
\psfrag{transv}[l][l][0.8]{$u_{z}\,$trans.}
\centering
\vspace{0.5cm}
   {   \subfloat{  \begin{tikzpicture}
        \node[anchor=south west,inner sep=0] (image) at (0,0) {\includegraphics[width = \textwidth]{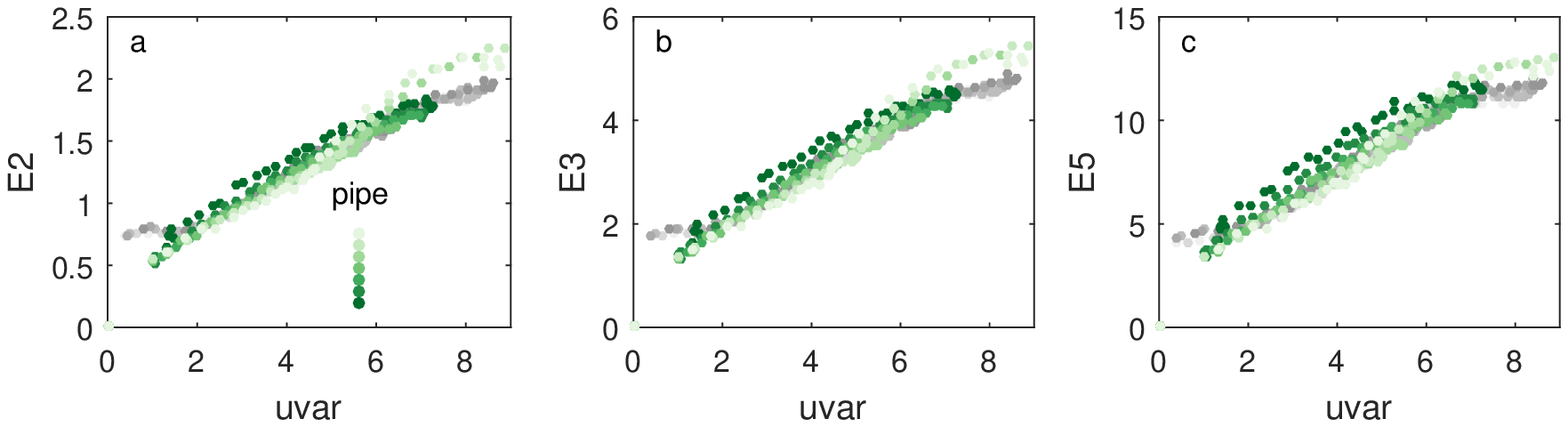}};
        \begin{scope}[x={(image.south east)},y={(image.north west)}]
                \node[inner sep=0,anchor=center] (note1) at (0.2,0.38) {\scalebox{0.8}{$Re_{\tau}$}};
                \node[inner sep=0,anchor=east] (note1) at (0.295,0.455) {\scalebox{0.8}{2000}};
                \node[inner sep=0,anchor=east] (note1) at (0.295,0.295) {\scalebox{0.8}{94400}};
                \draw [->] (0.2185,0.45) -- (0.2185,0.3);
                \draw [-] (0.235, 0.455) -- (0.242, 0.455);
                \draw [-] (0.235, 0.295) -- (0.242, 0.295);
        \end{scope}
    \end{tikzpicture}}}\\ 
{   \subfloat{ 
\psfrag{uvar}[c][c][1]{$\langle u_{\phi,x}^{\prime 2}\rangle^+, \, \langle u_z^{\prime 2}\rangle^+$}
 \begin{tikzpicture}
        \node[anchor=south west,inner sep=0] (image) at (0,0) {\includegraphics[width = \textwidth]{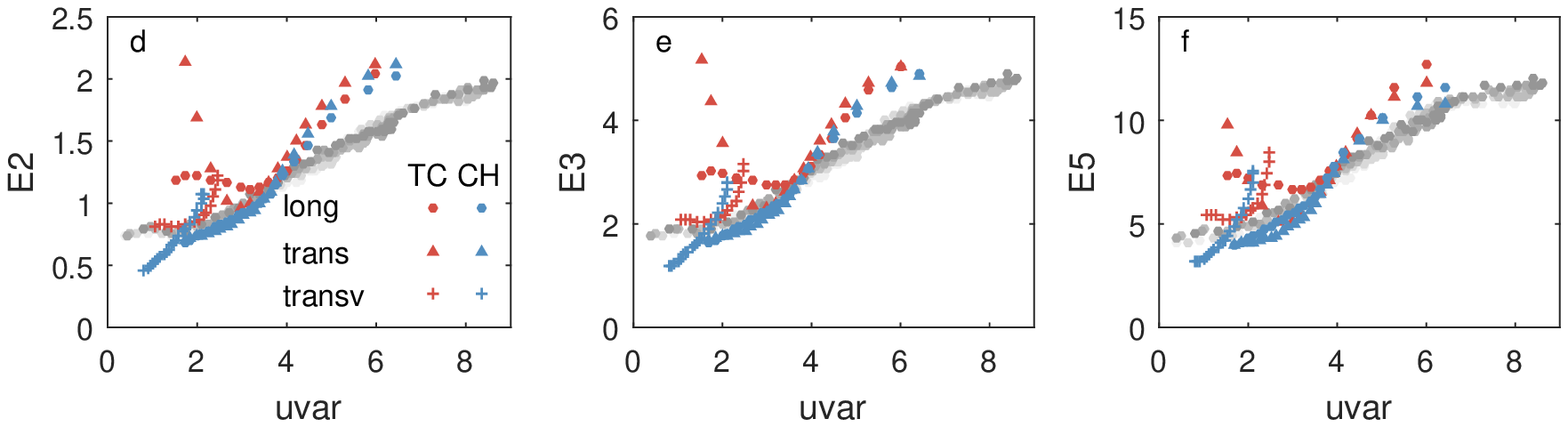}};
        \begin{scope}[x={(image.south east)},y={(image.north west)}]
        \end{scope}
    \end{tikzpicture}}}\\     
    \caption{ \label{fig:estar_univ} (a-c) Results for $E^*_{p,1}$ from the Superpipe data over a range $2000<Re_\tau <94400$ (shades of green) as a function of the streamwise variance $\langle u_x^{\prime 2}\rangle^+$. The legend in (a) also applies to (b-c). (d-f) Additive constant $E^*_{p,1}$ for channel and  TC flow for different directions and velocity components (cf. legend in (d)); results are shown for $30 \leq y^+ \lesssim 500$. 
    In all panels, the  TBL results from figure \ref{fig:estar}(d-f) serve as a reference and are plotted in shades of grey. 
      }
\end{figure}

We now return to the question of how the additive constants $E^*_{p,m}$ change across flow geometries. For pipe flow, the Superpipe data set of \citep{Hultmark2012} provides access to Reynolds numbers even exceeding those achieved in TBL experiments. In figure \ref{fig:estar_univ}(a-c), we compare results from the Superpipe to those of the TBL previously reported in figure \ref{fig:estar}(d-f). The pipe data are plotted for $y^+>10$  at the lowest three $Re_\tau$  ($Re_\tau\leq 5400$) but only for $y^+>300$ at higher $Re_\tau$ in order to stay clear of spatial resolution issues \citep[cf.][]{Smits2011}. In general, very good agreement is observed between the two geometries even at 10th order with possibly a slight trend of increasing $E^*_{p,1}$ with increasing $Re_\tau$ in the pipe. It is further noteworthy, that the pipe results for $Re_\tau\leq 5400$ `peel off' the TBL data at  $\langle u_x^{\prime 2}\rangle^+$-values corresponding to the vicinity of the inner peak, i.e.  $\langle u_x^{\prime 2}\rangle^+>6$. A comparable trend is not observed for higher $Re_\tau$ and it therefore appears plausible that the peel-off is a low-$Re_\tau$ effect rather than geometry related even though data very close to the inner peak (located at $y^+\approx 15$) are not available at the highest $Re_\tau$.  The fact that at nominally comparable $Re_\tau$ the low-$Re_\tau$ behaviour is different for the TBL might be related to the difficulty to define consistent outer length scales (and hence consistent values of $Re_\tau$) across different flow geometries \citep{Marusic2010key}.  

The results for the streamwise velocity component of the channel in figure \ref{fig:estar_univ}(d-f) (shown as filled blue symbols) display similar behaviour to the pipe at low Reynolds numbers. At lower values of $\langle u_x^{\prime 2}\rangle^+$ (further from the wall), the agreement with the TBL results is good but there are significant deviations when approaching the inner peak that start at slightly lower values of $\langle u_x^{\prime 2}\rangle^+$ in this case. It is further striking how closely the results for the longitudinal (circles) and transversal (triangles) directions agree. 

Values of $E^*_{p,1}$ corresponding to structure functions of $u_\phi$ in TC flow in both longitudinal and transversal directions (red filled symbols in figure \ref{fig:estar_univ}(d-f)) are observed to agree closely with the channel flow results closer to the wall ( higher values of  $\langle u_x^{\prime 2}\rangle^+$). In doing so, they also approach the TBL data around  $\langle u_x^{\prime 2}\rangle^+ \approx 4$ (equivalent to $y^+\approx 100$ cf. figure \ref{fig:meanvar}b). Unlike for the channel, both directions are seen to deviate upwards from the TBL results in the case of TC flow already at slightly lower values of  $\langle u_x^{\prime 2}\rangle^+$. At this point, also the agreement between longitudinal and transversal structure functions ceases. This anisotropy hints that this deviation from the behaviour of the other geometries may be related to the strong large scale roll modes (Taylor rolls) present in the bulk region of TC flow. It further remains open whether there is indeed a range of universality for $E^*_{p,1}$ that also encompasses TC flow at higher $Re_\tau$, where a more pronounced logarithmic region is present.

For completeness, we also report the $E^*_{p,1}$ values for the spanwise velocity component $u_z$ in the transversal direction in figures \ref{fig:estar_univ}(d-f) for channel and TC flow plotted vs. $\langle u_z^{\prime 2}\rangle^+ $. However, due to the lack of reference data at high $Re_\tau$ for this case, not much more can be stated than that - similar to the observations for $u_{x,\phi}$ - also the results for $u_z$ differ at lower values of $\langle u_z^{\prime 2}\rangle^+ $.

\section{Summary, conclusions and outlook} \label{sec:CONC}
Various aspects of the analysis in the spirit of the extended self-similarity (ESS) framework applied to the energy containing range of structure functions in wall bounded flows were considered in this work. A simple model based on the attached eddy hypothesis proved helpful in providing a theoretical underpinning to the key features of the analysis in the spirit of the ESS hypothesis. Namely, it was shown that when considering relative scalings, the self similarity assumption for the eddies is relaxed. In the relative form, the scaling relationships are therefore more robust against viscous effects or large scale perturbations, leading to more robust scaling regions that are also discernible at low values of $Re_\tau$, which are accessible through direct numerical simulations.
Further, different logarithmic slopes in longitudinal and transversal structure functions were seen to exist in the present data as well as previous studies. The attached-eddy model explains this observation in terms of a tendency of the eddies to align in the streamwise direction, which is consistent with the well-documented presence of long streamwise structures in the logarithmic region of wall-bounded flows. This effect, too, cancels out in the ESS framework, conforming with the observations here (for TC and channel flow) and in \citet{deSilva2017} (for TBL) that the ratios $D_p/D_m$ for the streamwise velocity component are approximately the same in both directions. The model further predicts an analogous behaviour, which could be confirmed at least for the transversal direction of TC and channel flow in this work. Further work at higher $Re_\tau$ will be needed to see if this holds also for $\SFnew[x]{x}{p}$ in canonical wall-bounded flows.

Focussing on TC flow, we found generally good agreement with ESS results in other geometries. This applies to the mere existence (at least in the ESS-framework) of a logarithmic scaling region, which had not been established before in TC flow, but also the magnitude of the ratios $D_p/D_m$. This is remarkable, especially in view of the fact that the presence of the large scale Taylor rolls significantly alters the appearance of $\SFTnew[z]{x}{p}$ in the direct representation in figures \ref{fig:udir}(d-f) compared to the corresponding channel flow result. The only exception to this agreement is $\SFnew[z]{z}{p}$, where the data distinctively deviates from the linear relationship predicted in the ESS form. For this configuration, there is a direct contribution of the Taylor rolls to the structure function, which, unlike for $\SFTnew[x]{z}{p}$, does not approximately cancel out in the mean when taking the velocity difference between two points. From an attached eddy-perspective, these results can be interpreted as follows: The boundary layer in TC flow consists of wall attached structures which are modulated by the super-imposed Taylor rolls (TR). For $\SFTnew[z]{x}{p}$, there is no superposition from the TRs since their  $u_\phi$ component is zero. Nevertheless, their modulation of the attached eddies leads to pronounced low and high velocity streaks but this effect gets cancelled out when taking the ratios in the ESS framework. However, for structure functions of $u_z$, the TR directly contribute such that additional summands enter in (\ref{eq:add}) or (\ref{eq:add2}), which do not adhere to the hierarchical organisation of the wall-attached eddies. Unless their contribution drops out in the mean when taking the velocity difference (as is the case for $\SFTnew[x]{z}{p}$), these will alter the distribution of $\Delta u_z$ in which case a scaling according to (\ref{eq:du2du-ESS}) can no longer be expected. The latter applies for $\SFnew[z]{z}{p}$ and is consistent with the failure to observe ESS-scaling there.

In another aspect of this work, we addressed the behaviour of the additive constants $E^*_{p,m}$, which constitutes the only other constant in the ESS form (\ref{eq:ESS}). Our analysis for $u_x$ in the TBL revealed that $E^*_{p,m}$ exhibits significant dependencies on the wall-normal position (increasing with increasing $y_0^+$) and the Reynolds number (increasing with increasing $Re_\tau$). However, the data collapses when plotted versus $\var[x]^+$ instead of $y_0^+$ --- a behaviour that can be explained by considering (\ref{eq:ESS}) in the limit of $s_i \to \infty$. Doing so  yields a linear dependence of $E^*_{p,m}$ on $\var[x]^+$ in the logarithmic region with slopes that are largely consistent with the data. Further, $E^*_{p,m}$ was seen to display universal behaviour in this form at least for pipe flow, where data at large $Re_\tau$ are available. The present low-$Re_\tau$ datasets do not allow for a conclusive judgement as to whether this universality extends to channel flow or potentially even TC. While this appears likely for the former, there are indications that, presumably due to the presence of the TR, there are significant deviations from the TBL results in the outer regions of the boundary layer in TC flow. Future TC data at higher $Re_\tau$ will help to shed light on this as well as on the behaviour of $E^*_{p,m}$ for $u_z$, which could not be assessed from the present data.

\section*{Acknowledgements}
 The authors would like to express their gratitude to Prof. Alexander Smits for sharing the Superpipe data. We further would like to thank Juan C. del Alamo for making the channel simulations available. The TC data was obtained using the PRACE project 2013091966 resource CURIE based in France at Genci/CEA. Continuous financial support by the Australian Research Council, Dutch NWO and from ERC is gratefully acknowledged. DK acknowledges financial support by the University of Melbourne through the McKenzie fellowship. ROM was supported in part by the National Science Foundation under Grant No. PHY11-25915.

\bibliographystyle{jfm}
\bibliography{jfm_TC}

\end{document}